\documentclass[a4paper,fleqn]{cas-dc-r1}

\usepackage[numbers]{natbib}
\usepackage{caption}
\usepackage{algorithm}
\usepackage{algpseudocode}
\usepackage{pifont}

\def\tsc#1{\csdef{#1}{\textsc{\lowercase{#1}}\xspace}}
\tsc{WGM}
\tsc{QE}
\tsc{EP}
\tsc{PMS}
\tsc{BEC}
\tsc{DE}

\usepackage{enumitem}

\setlist[itemize]{itemsep=0pt, topsep=0pt, partopsep=0pt, parsep=0pt}

\begin{document}
\let\WriteBookmarks\relax
\def\floatpagepagefraction{1}
\def\textpagefraction{.001}
\shorttitle{}
\shortauthors{Lin Meng et~al.}

\title [mode = title]{MemX: Memory-Aware Automatic Parallelism for Cross-Vendor Heterogeneous GPU Training}                      

\author[1,2]{Lin Meng}

\affiliation[1]{organization={Institute of Computing Technology, Chinese Academy of Sciences},
                city={Beijing},
                postcode={100190}, 
                country={China}}

\author[1]{Yuzhong Sun}[orcid=0000-0002-0362-4420]
\ead{yuzhongsun@ict.ac.cn}
\cormark[1]

\affiliation[2]{organization={University of Chinese Academy of Sciences},
                postcode={101408}, 
                country={China}}

\author[3]{Jie Zhu}

\affiliation[3]{organization={School of Computer Science, Nanjing University of Posts \& Telecommunications},
                postcode={210023}, 
                country={China}}

\cortext[cor1]{Corresponding author at: Institute of Computing Technology, Chinese Academy of Sciences, 100190, China.}

\begin{abstract}
Cross-vendor heterogeneous GPU clusters are increasingly used for large-scale model training. However, their GPUs differ substantially in compute power, memory capacity, and interconnect characteristics. Existing automatic parallelism training systems usually minimize iteration time while treating memory as a hard constraint, almost failing in resolving the compute--memory asymmetry of cross-vendor clusters regardless correlation between memory and throughput. In this paper, we present MemX, a new memory-aware automatic parallelism system that jointly optimizes iteration time, peak memory usage, and cross-GPU memory balance. MemX consists of a weighted-product objective function, heterogeneous recomputation strategies, a minimal-set profiler, interpolation-based data augmentation, a dual-output XGBoost predictor, and a hash-signature prediction cache, in order to accelerate search of large-scale auto-parallel best configurations at lower cost. We implement MemX on Megatron-LM and evaluate it against four automatic parallelism baselines on two heterogeneous clusters. MemX achieves the lowest prediction Mean Absolute Percentage Error (MAPE) of 12.17\% and is the only method that identifies the real Top-1 configuration under both time-based and memory-balance rankings. In terms of search cost and search space, MemX searches 31,240 configurations in 26.5 seconds, while an end-to-end profiling baseline is extrapolated to require about 17 days at the same scale. The simulation study that extends the measured 20-GPU data to 40-GPU and 60-GPU clusters further illustrates the effectiveness of MemX at larger scales.
\end{abstract}


\maketitle

\section{Introduction}
The rapid growth of large language models (LLMs), such as GPT-5 and GLM-5
\citep{singh2025openai, zeng2026glm}, has created a huge increase in computational demand. Training these models, which comprise hundreds of billions to trillions of parameters, typically requires thousands of high-performance GPUs operating continuously for months \citep{touvron2023llama,touvron2023llama2}. Current approaches generally deploy large-scale homogeneous clusters of NVIDIA GPUs, benefiting from their mature software ecosystem. However, this approach faces several limitations: geopolitical tensions and supply chain shortages have made building large-scale homogeneous clusters increasingly difficult, while the high costs and energy consumption of cutting-edge accelerators compel organizations to seek alternative hardware solutions.

Therefore, the emergence of cross-vendor heterogeneous GPU clusters has become an inevitable trend in large-scale model training. It is essential to distinguish between traditional intra-vendor generational heterogeneity (e.g., mixing NVIDIA V100, A100, and H100) and cross-vendor architectural heterogeneity (e.g., AMD, Huawei Ascend, Biren) \citep{MI300X, biv150, Gaudi}, where GPUs differ in instruction sets, memory architectures, and interconnect technologies. The latter presents fundamentally distinct system challenges that existing training frameworks fail to handle effectively.

To bridge this hardware diversity and achieve efficient training, automatic parallelism strategies capable of adapting to heterogeneous GPU capabilities are required. Current automatic parallelism systems for intra-vendor heterogeneous GPUs commonly employ pipeline parallelism (PP) \citep{huang2018gpipe, narayanan2019pipedream} partitioning as the core mechanism for load balancing: allocating layers proportionally to device compute capacity \citep{li2022amp, um2024metis, yanhexiscale, hwang2024FaSOP}, where high-performance GPUs assume more layers and low-performance devices assume fewer. This strategy performs well in intra-vendor generational heterogeneity scenarios because device performance is typically positively correlated with memory capacity. For instance, compared with the NVIDIA V100 (125 TFLOPS/32GB), the NVIDIA A100 (312 TFLOPS/80GB) exhibits exhibits a 2.5× increase in both computing power and memory, with roughly matching ratios.

\begin{table}[width=\linewidth,cols=5,pos=h]
\centering
\caption{Heterogeneous PP partitioning failure.}
\label{tab:motivating}
\begin{tabular}{p{19mm}p{7mm}p{10mm}p{13mm}p{13mm}}
\toprule
\textbf{Strategy} & \textbf{Layers} & \textbf{TFLOPS} & \textbf{Vendor X} & \textbf{Vendor Y} \\
 & \textbf{(X:Y)} & & \textbf{Mem\%} & \textbf{Mem\%} \\
\midrule
Compute & 21:11 & $OOM^{*}$ & $>$100 & — \\
Memory & 8:24 & 55 & 88 & 78 \\
\midrule
Vendor X only & 8:0 & 70 & 88 & — \\
Vendor Y only & 0:24 & 86 & — & 78 \\
\bottomrule
\multicolumn{5}{l}{\footnotesize $^{*}$ Out-of-Memory.}
\end{tabular}
\end{table}

However, the memory-performance asymmetry in cross-vendor heterogeneous scenarios renders such a partitioning mechanism ineffective. We validate this observation on a employed cluster with Vendor X ($\sim$300 TFLOPS/32GB) and Vendor Y ($\sim$256 TFLOPS/64GB), using partitioning ratios derived from the effective capabilities measured on real training workloads (i.e., LLaMa 2-7B with 32 layers) rather than the nominal specifications. Although the nominal peak-compute ratio is only 300:256 ($\approx$1.17:1), differences in kernel efficiency and software-stack maturity give X about 1.9$\times$ the effective training throughput of Y, i.e., a 21:11 compute ratio; likewise, despite the 32:64 (1:2) nominal memory ratio, the same workload exhibits different memory footprints on the two vendors' software stacks, so the memory-feasible assignment is 8:24 (1:3) rather than 11:21. As shown in Table~\ref{tab:motivating}, the 21:11 computing-proportional partition exceeds X's 32GB memory triggering Out-of-Memory, while the 8:24 memory-proportional partition runs but achieves merely 55 TFLOPS, far below the homogeneous baselines of 70 (X-only) and 86 (Y-only) TFLOPS. The culprit is a stage-time mismatch in PP: at these standalone rates, X computes its 8 layers in only (8/70)/(24/86) $\approx$ 41\% of the time which Y takes for its 24 layers, idling for about 59\% of each pipeline cycle, while Y becomes the bottleneck despite 78\% memory utilization; overall, the pipeline harvests only 35\% of the two devices' combined 156 TFLOPS (70$+$86), wasting about 65\% (101 TFLOPS) of the available computing power. As the computing/memory partitioning mechanisms optimize only a single dimension of load balancing, they are inherently incapable of handling the dual constraints of computing power and memory simultaneously \citep{li2022amp, um2024metis, yanhexiscale, hwang2024FaSOP}.

Beyond the memory-performance asymmetry, cross-vendor heterogeneity also introduces severe challenges in performance modeling. Automatic parallelism requires finding the optimal solution within a vast configuration space \citep{jia2019beyond, miao2022galvatron} determined by multiple dimensions, including parallelization strategies (data/tensor/pipeline parallelism \citep{narayanan2021efficient}, etc.), model partitioning strategies, recomputation strategies \citep{chen2016training, korthikanti2023reducing}, and other training parameters (such as micro-batch size and number of attention heads), resulting in search state explosion. For example, LLaMA 2/3 \citep{touvron2023llama2, grattafiori2024llama} layers range from 32 upto 80. For a 32-layer Llama 2 model on only 16 GPUs, with 3 parallelization modes, 3 recomputation strategies, and 5 categories of model parameters (each with about 3 values), the configuration space already reaches the order of $10^8$, while the 80 layers of LLaMA 3 has $10^9$ configuration space. Exhaustively profiling this space would require decades, rendering it clearly infeasible.

Existing automatic parallelism systems typically employ performance models based on sparse sampling or search pruning to reduce overhead \citep{zhu2025mist, um2024metis, strati2025sailor, li2025malleus}. However, the limited profiling budget conflicts with the discrete-continuous hybrid feature space in cross-vendor scenarios (GPU types, batch sizes, sequence lengths, etc.), where sampling sparsity leads to significant degradation in prediction accuracy. Furthermore, current approaches cannot predict performance for configurations outside the sampled range. When new GPUs are added, the entire profiling process must be repeated, resulting in low deployment efficiency and poor adaptability to expanding heterogeneous clusters.

To address these challenges, we present MemX, a memory-aware automatic parallelism system that jointly optimizes iteration time, peak memory usage, and cross-GPU memory balance for cross-vendor heterogeneous GPU clusters. Beyond avoiding out-of-memory failures, such memory-balanced configurations are more robust to runtime memory fluctuations, easier to schedule in shared clusters, and more portable across heterogeneous GPU types. The core contributions of MemX are threefold.

\begin{itemize}[]
    \item \textbf{Memory-aware heterogeneous parallelism search.} To tackle the load imbalance caused by memory-performance asymmetry, MemX explicitly incorporates peak memory usage as an optimization objective alongside training throughput. The key innovation is support for heterogeneous recomputation strategies: memory-constrained GPU devices adopt aggressive recomputation to trade time for space, while memory-abundant devices use conservative recomputation to trade space for time, thereby minimizing total training time under memory constraints.
    \item \textbf{Minimal-set profiler with data augmentation.} To address the tension between exploding configuration space and profiling overhead, MemX designs a minimal-set profiler that identifies critical parameter subsets for sparse sampling, reducing profiling overhead by orders of magnitude; augments training data based on linear interpolation assumptions; and employs a machine learning model to achieve millisecond-level performance prediction with a mean absolute percentage error (MAPE) of 12.17\%.
    \item \textbf{Real-system validation.} We implement MemX\footnote{https://github.com/meng980626/MemX.} based on Megatron-LM \citep{narayanan2021efficient, shoeybi2019megatron} and evaluate it on two real cross-vendor GPU clusters. Experiments on LLaMA2-7B/ 13B/70B and LLaMA3-8B/70B models show that MemX reduces mean MAPE by 18.4\% compared with the strongest baseline, reduces profiling measurements by a factor of about 411 in the five-dimensional example, searches 160$\times$ as many configurations as AMP, and is over 1{,}400$\times$ faster at the 2{,}000-configuration scale.
\end{itemize}


\section{Background}

This section reviews the background that motivates MemX. Section 2.1 explains why cross-vendor heterogeneous GPU clusters are becoming inevitable and quantifies the compute--memory asymmetry that distinguishes them from homogeneous settings. Section 2.2 surveys automatic parallelism systems designed for heterogeneous clusters and analyzes their limitations under this asymmetry. Section 2.3 reviews performance modeling approaches and discusses the tension between modeling accuracy and profiling cost, which motivates MemX's lightweight modeling design in Section 3.2.

\subsection{Inevitability of Heterogeneous Trends}

The ever-increasing computational demands of large model training, coupled with constrained supply from single-vendor GPUs, have made cross-vendor heterogeneous clusters an inevitable reality. Accelerators such as AMD MI300X \citep{MI300X}, Intel Gaudi 3 \citep{Gaudi}, Huawei Ascend 910B, Biren BR100 series, and ILUVATAR BI-V150 \citep{biv150} have gradually achieved large-scale deployment capabilities, collectively challenging NVIDIA's dominance. These GPUs vary widely in key metrics including peak compute power, memory capacity, and interconnect bandwidth, compelling training frameworks to adapt to cross-vendor mixed deployments.

\begin{figure}
	\centering
	\includegraphics[width=\linewidth]{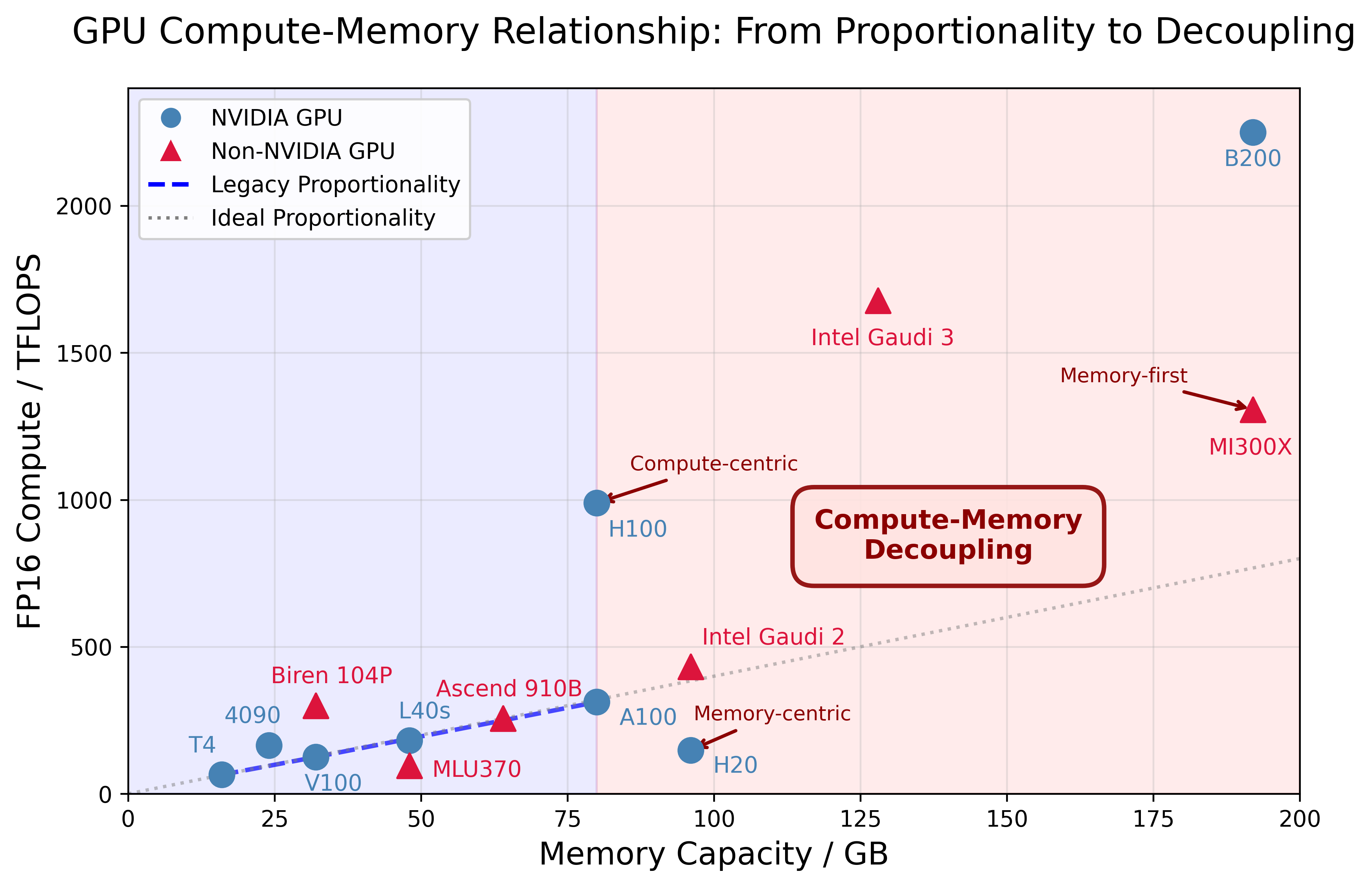}
    \vspace{-20pt}
	\caption{Peak FP16 compute versus memory capacity of mainstream GPUs. Legacy NVIDIA GPUs (T4, V100, A100) follow a stable compute-to-memory proportionality (dashed line), while newer and non-NVIDIA devices deviate in both directions: compute-centric (H100, 12.4:1) or memory-centric (H20, 1.5:1), marking a decoupling of compute and memory in emerging heterogeneous clusters.}
	\label{FIG:gpu_het}
\end{figure}

Figure \ref{FIG:gpu_het} illustrates the distribution of peak computing power and memory capacity across mainstream GPUs. Early-generation NVIDIA GPUs (T4, V100, A100) exhibit a clear positive correlation, with compute-to-memory ratios stabilizing at approximately 4:1. However, newer GPUs deviate significantly from this pattern: the NVIDIA H100 (989 TFLOPS/80GB) achieves substantial compute gains without proportional memory increases, yielding a compute-to-memory ratio of 12.4; consumer-grade GPUs such as the RTX 3090 (142 TFLOPS/24GB) and RTX 4090 (165 TFLOPS/24GB) demonstrate high-compute/low-memory characteristics. Inference-optimized GPUs like the H20 (148 TFLOPS/96GB), with compute reduced to 15\% of H100 yet retaining 96GB memory, exhibit an exceptionally low compute-to-memory ratio of merely 1.5. Cross-vendor devices further exacerbate this non-monotonicity: the Biren BR104P ($\sim$300 TFLOPS/32GB) delivers A100-comparable compute with only 40\% of the memory, while the Huawei Ascend 910B ($\sim$256 TFLOPS/64GB) offers lower compute yet double the memory. This decoupling of compute and memory capacities becomes the main difference between emerging cross-vendor heterogeneous clusters and traditional homogeneous deployments. Recent systems have begun to automate distributed training over heterogeneous, geo-distributed, or dynamically changing resources \citep{strati2025sailor, zong2025training, li2025malleus, liang2026harp, wang2026autohaap}, further highlighting the practical importance of heterogeneity-aware parallelization.

\subsection{Heterogeneous Automatic Parallelism}

Distributed training frameworks combine data, tensor, and pipeline parallelism to scale large models across GPU clusters \citep{shoeybi2019megatron, huang2018gpipe, narayanan2019pipedream}, and automatic parallelism systems further search for model partitioning and device placement strategies instead of relying on manual configuration \citep{zheng2022alpa, miao2022galvatron}. For heterogeneous clusters, AMP explores tensor, pipeline, data parallelism, and micro-batch size through heterogeneity-aware strategy search \citep{li2022amp}; Metis extends this line with device-group partitioning for heterogeneous GPUs \citep{um2024metis}; HexiScale uses graph partitioning and analytical cost modeling \citep{yanhexiscale}; and FaSOP assigns pipeline stages through a Minmax-Stage layer-assignment algorithm \citep{hwang2024FaSOP}. More recent systems extend automatic parallelism to broader heterogeneous and dynamic settings, including geo-distributed or dynamically available resources \citep{strati2025sailor, zong2025training}, straggler-aware replanning \citep{li2025malleus}, commodity-server tensor parallelism \citep{li2025oases}, operator-level asymmetric partitioning \citep{wang2026autohaap}, and fine-grained heterogeneous cluster planning \citep{liang2026harp}. HPMD further adapts GPU allocation and hybrid parallelism under changing batch sizes \citep{yun2026hpmd}.

These systems commonly optimize iteration time, monetary cost, or load balance, and usually handle memory as a feasibility constraint or a component of the cost model. Although several recent systems model memory footprint or incorporate memory-saving techniques \citep{zhu2025mist, strati2025sailor, li2025malleus, li2025oases, wang2026autohaap}, they generally do not jointly optimize iteration time, peak memory usage, and cross-GPU memory balance for cross-vendor heterogeneous clusters.

Although those systems employ pipeline parallelism (PP) slicing to achieve load balancing by allocating layers proportionally to device compute power, such strategies fail to generalize across all heterogeneous scenarios. Their efficacy relies on an implicit assumption: computing power ratios align with memory requirement ratios. In cross-vendor heterogeneous scenarios, the decoupling of compute and memory renders single-dimensional slicing strategies incapable of satisfying dual constraints simultaneously. Strictly following compute proportions leads to memory bottlenecks, while accommodating memory capacities results in compute underutilization. This dilemma exposes the fundamental limitation of traditional approaches in handling asymmetric heterogeneity.

Furthermore, existing systems typically adopt global uniform policies for training parameters such as recomputation, including fully enabled, fully disabled, or layer-wise uniform configurations, lacking adaptive capabilities for heterogeneous memory capacities. This rigid design leads to significant performance degradation in cross-vendor scenarios: memory-constrained devices face overflow risks when forced to adopt conservative policies, while memory-abundant devices waste resources when subjected to aggressive policies. The inability of uniform strategies to accommodate differentiated device requirements becomes another critical bottleneck restricting training efficiency.

Beyond guaranteeing feasibility, jointly optimizing memory balance yields practical benefits that a time-only objective cannot provide. The two memory terms in the objective correspond to two distinct risks. High memory utilization (M) exposes a configuration to out-of-memory failures. Measured memory usage is not a static quantity: allocator fragmentation, variable-length batches, and transient communication buffers all introduce fluctuations over a long training run, and a configuration operating close to device capacity can be aborted by a single transient spike, wasting days of computation. Configurations with moderate utilization are robust to such fluctuations, and they are also friendlier to cluster scheduling: a modest and predictable per-device footprint is easier to place and bin-pack, and can be more readily co-located with other tenants in shared clusters.

High memory variance (V) across devices creates a different risk: the bottleneck GPU device. The feasibility and stability of the entire configuration become hostage to the single most-loaded device, even when the average utilization is low. Configurations with uniform memory headroom avoid this risk in two ways. First, they facilitate subsequent hyper-parameter exploration. Adjustments such as enlarging the micro-batch size or the sequence length introduce additional memory that must be absorbed on every device, which is only possible when the headroom is uniform; researchers can therefore vary hyper-parameters around a known-good configuration without re-running the full configuration search. Second, they improve portability across hardware. A configuration with low and balanced utilization carries a compatibility margin that allows it to fit the different GPU types of a heterogeneous cluster, eliminating the cost of re-searching whenever the deployment target changes. These considerations motivate treating memory balance as a first-class optimization objective rather than a binary constraint.


\subsection{Performance Modeling}

Performance modeling serves as a core component of automatic parallelism systems, responsible for predicting the execution efficiency of different parallel configurations to provide decision support for search algorithms. Existing modeling approaches are dominated by analytical and measurement-based models, while recent work further introduces simulation-based and learning-based variants.

Some prior works construct analytical cost models by parsing static features of computational graphs, estimating FLOPs, communication volume, and memory footprint to predict execution performance \citep{jia2018exploring, wang2019supporting, song2019hypar, jia2019beyond, song2020accpar, cai2021tensoropt, zheng2022alpa, zhang2024hap, liu2023colossal}. HGTrainer relies on such analytical computation, communication, and memory models for heterogeneous geo-distributed clusters \citep{zong2025training}, and Mist further uses symbolic expressions to avoid repeated simulation during configuration search \citep{zhu2025mist}. The advantage of such methods lies in low runtime overhead, enabling rapid evaluation of different configurations without actual execution. However, their limitations are equally significant: they struggle to capture dynamic characteristics during actual execution, including GPU caching behavior, compute utilization fluctuations, kernel launch overhead, and complex hardware interactions such as vendor driver optimization differences. In cross-vendor heterogeneous scenarios, microarchitectural differences and software stack characteristics across vendors further amplify the impact of these unmodeled factors, leading to substantial prediction errors.

Other works have shifted toward execution time sampling, recording durations from short iterations or operators on real hardware and constructing cost models to predict performance of complete configurations \citep{narayanan2019pipedream, narayanan2021memory, fan2021dapple, miao2022galvatron, liu2024aceso, qi2026hetauto, zhang2025poplar}. Measurement granularity presents a trade-off. Iteration-level sampling executes full forward and backward propagations. It offers high accuracy but requires longer measurement time. Operator-level sampling targets individual operators, which is faster but requires compositional estimation and may introduce cumulative errors. Both approaches share a common limitation: profiling overhead grows explosively with parameter combinations. Different parallelization strategies and training hyperparameters such as batch sizes, sequence lengths, and recomputation configurations require independent sampling. The configuration space scale directly determines profiling cost. Although sparse sampling can reduce such overhead, balancing modeling accuracy against profiling cost remains a significant challenge.

Another line of work instead measures execution time on real hardware \citep{narayanan2019pipedream, narayanan2021memory, fan2021dapple, miao2022galvatron, liu2024aceso, qi2026hetauto, zhang2025poplar}. Short iterations or individual operators are timed, and the measurements are used to fit a cost model that predicts the performance of complete configurations. Measurement granularity presents a trade-off. Iteration-level sampling runs full forward and backward propagation; it is accurate but slow. Operator-level sampling is faster, but it requires compositional estimation and may accumulate errors. Both approaches share a common limitation: profiling overhead grows explosively with the number of parameter combinations. Every combination of parallelization strategy and training hyperparameters must be sampled independently, so the profiling cost scales directly with the size of the configuration space. Sparse sampling reduces this overhead, but balancing modeling accuracy against profiling cost remains a significant challenge.

Taking LLaMA2-7B (N=32 layers) training on 16 GPUs as an example, and constraining pipeline parallelism degree to $PP\leq4$, the configuration space is determined by four dimensions: 15 base parallel strategies satisfying TP×PP×DP=16 with each dimension as a power of two; $\binom{N+PP-1}{PP-1}$ layer partitioning schemes for PP=4; $r^{PP}$ recomputation combinations from r strategies per stage; and $x^y$ configurations from x model hyperparameters (batch size, sequence length, etc.) each with y possible values. The total configuration space under $PP\leq4$ constraint reaches approximately $10^8$ (billion-scale). Complete profiling would require decades, while existing sparse sampling methods lack effective prior guidance in cross-vendor scenarios, making the trade-off between modeling accuracy and profiling overhead a fundamental challenge.

Beyond these two dominant lines, simulation-based approaches model runtime and memory behavior through resource-level or graph-based simulators \citep{strati2025sailor, guo2025adaptis, wu2025rethinking}; hybrid systems combine profiling with cost estimation or cached intermediate results \citep{li2025malleus, wang2026autohaap, liang2026harp}; AutoScout reduces profiling cost by prioritizing high-impact configuration features and adaptively combining simulators with different fidelities \citep{shong2026autoscout}; and learning-based solvers such as NNPiper and APSS train neural strategy generators offline using cost-model feedback \citep{ou2025neural, ou2025efficient}. MemX adopts a hybrid design: a minimal-set profiler measures only high-impact parameter combinations, validated interpolation augments the training set, and a dual-output XGBoost predictor estimates both time and memory, while a hash-signature cache avoids repeated predictor calls when candidate configurations share performance-critical substructures.

\subsection{Memory Optimization and Pipeline Scheduling} 

Activation recomputation reduces memory footprint by discarding intermediate activations and recomputing them during backpropagation \citep{chen2016training}, while selective recomputation preserves a subset of expensive activations to balance time and space \citep{korthikanti2023reducing}. Recent systems increasingly co-optimize memory-saving techniques with parallelization: Mist jointly tunes activation checkpointing, redundancy elimination, offloading, and parallelism \citep{zhu2025mist}; AutoHAAP combines memory-aware initialization with activation recomputation selection \citep{wang2026autohaap}; and Oases models overlapped communication and computation under tensor model parallelism \citep{li2025oases}. Pipeline schedules such as 1F1B, interleaved 1F1B, and Chimera reduce pipeline bubbles through different micro-batch execution orders \citep{narayanan2019pipedream, shoeybi2019megatron, li2021chimera}. OctoPipe further co-optimizes partitioning, placement, and scheduling through a graph-based pipeline simulator \citep{guo2025adaptis}, while HARP derives a heterogeneity-aware 1F1B variant to hide cross-cluster communication latency \citep{liang2026harp}. MemX differs in assigning recomputation strategies at pipeline-stage granularity according to device memory and compute characteristics, and jointly evaluating scheduling, parallelism, recomputation, and memory balance under a unified objective.

\begin{figure}
  \centering
  \includegraphics[width=\linewidth]{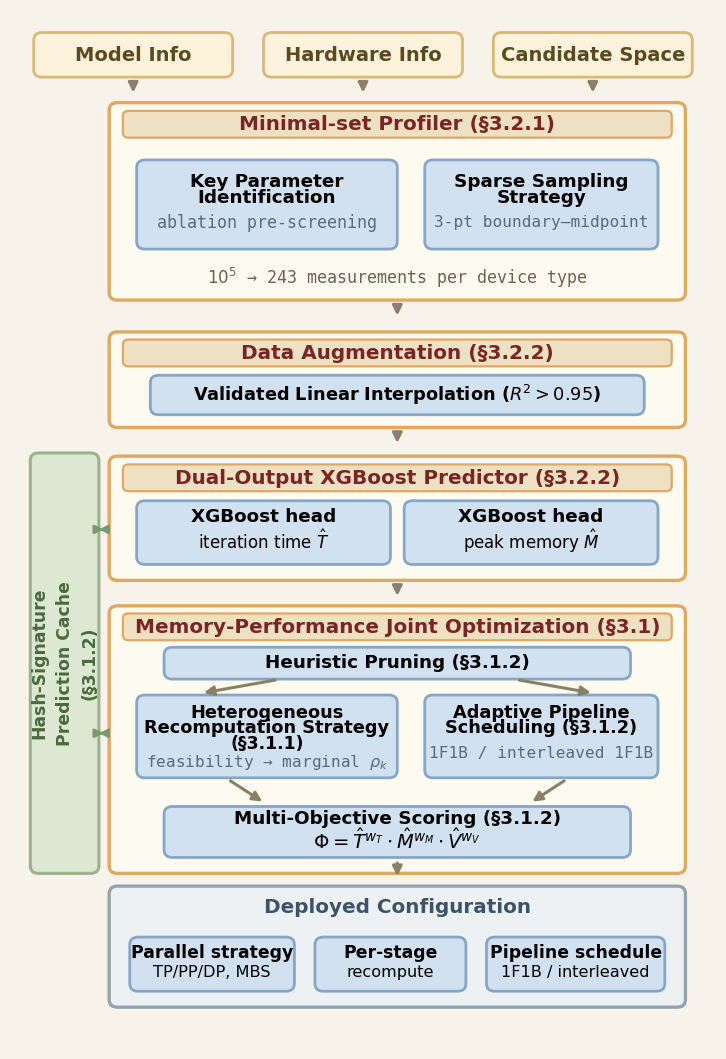}
  \caption{Overview of MemX. In the offline phase (Section~3.2), a minimal-set
  profiler measures a small set of high-impact parameter combinations on
  representative devices, validated interpolation augments the measurements, and
  a dual-output XGBoost predictor is trained to estimate iteration time and peak
  memory in a single inference call. In the online phase (Section~3.1), MemX
  enumerates the pruned candidate space, evaluates each configuration under the
  weighted-product objective $\Phi$ with the help of a hash-signature prediction
  cache, and outputs the deployed configuration together with its per-stage
  recomputation policy and pipeline schedule.}
  \label{fig:overview}
\end{figure}

\section{Method}

Figure~\ref{fig:overview} gives an overview of MemX, and Algorithm~\ref{alg:memx} summarizes the end-to-end procedure. MemX operates in two phases: an offline phase that builds lightweight performance models from a minimal set of real measurements (Section 3.2), and an online phase that searches the pruned configuration space under the memory--performance joint objective (Section 3.1). We first describe the search algorithm and its objective, and then the performance modeling pipeline that makes the search affordable.

\begin{algorithm}[t]
\caption{MemX configuration search}
\label{alg:memx}
\begin{algorithmic}[1]
\Require model $\mathcal{M}$ ($N$ layers, hidden size, heads), cluster $\mathcal{C}=\{(C_k,M_k)\}_{k=1}^{K}$, weight profile $(w_T,w_M,w_V)$
\Ensure deployed configuration $\mathbf{x}^*$
\State $\mathcal{S} \gets$ candidate space (TP, PP, DP, MBS, recompute, hyper-parameters) \Comment{\S 3.1.2}
\State $\mathcal{S} \gets$ \Call{HeuristicPrune}{$\mathcal{S}$} \Comment{\S 3.1.2}
\State $P \gets$ \Call{HighImpactParams}{$\mathcal{M},\mathcal{C}$} \Comment{\S 3.2.1}
\State $D \gets$ \Call{ThreePointSample}{$P$} \Comment{\S 3.2.1}
\State $D \gets D \cup$ \Call{Interpolate}{$D$} \Comment{\S 3.2.2}
\State $f \gets$ \Call{TrainDualXGBoost}{$D$} \Comment{\S 3.2.2}
\State $\mathrm{cache} \gets \emptyset$; \ $\Phi^* \gets \infty$; \ $\mathbf{x}^* \gets \emptyset$
\For{$\mathbf{x} \in \mathcal{S}$}
  \State $\sigma \gets$ \Call{HashSignature}{$\mathbf{x}$} \Comment{\S 3.1.2}
  \If{$\sigma \in \mathrm{cache}$}
    \State $(\hat{T},\hat{M}) \gets \mathrm{cache}[\sigma]$
  \Else
    \State $(\hat{T},\hat{M}) \gets f(\mathbf{x})$; \ $\mathrm{cache}[\sigma] \gets (\hat{T},\hat{M})$
  \EndIf
  \State $\mathbf{r} \gets$ \Call{RecomputePolicy}{$\mathbf{x},\mathcal{C}$} \Comment{\S 3.1.1}
  \State $s \gets$ \Call{SelectSchedule}{$\mathbf{x}$} \Comment{\S 3.1.2}
  \State $\hat{V} \gets$ per-device memory variance \Comment{\S 3.1.2}
  \State $\Phi \gets (\hat{T}/T_{\mathrm{ref}})^{w_T}\cdot(\hat{M}/M_{\mathrm{ref}})^{w_M}\cdot((\hat{V}+\epsilon_V)/(V_{\mathrm{ref}}+\epsilon_V))^{w_V}$ \Comment{\S 3.1.2}
  \If{$\Phi < \Phi^*$}
    \State $\Phi^* \gets \Phi$; \ $\mathbf{x}^* \gets (\mathbf{x},\mathbf{r},s)$
  \EndIf
\EndFor
\State \Return $\mathbf{x}^*$
\end{algorithmic}
\end{algorithm}

\subsection{Memory-Performance Joint Optimization Search Algorithm}
\subsubsection{Heterogeneous Recomputation Strategy}

Most existing automatic parallelism systems apply globally uniform recomputation policies or treat memory-saving techniques as separate from parallelism search. Recent systems have started to combine parallelism search with activation recomputation or other memory-saving techniques \citep{zhu2025mist, wang2026autohaap, li2025oases}. However, these designs primarily target memory footprint reduction or communication-computation overlap, rather than assigning heterogeneous recomputation policies to explicitly balance memory headroom across devices. Therefore, MemX allows differentiated recomputation strategies across pipeline stages (typically on different GPU types), enabling adaptive time-space tradeoffs.

The recomputation strategy is divided into three levels:

\begin{itemize}[]
    \item \textbf{Full Recomputation.} Only the input activations of each layer are retained; all intermediate variables during forward propagation (including attention matrices, Softmax outputs, Dropout masks, etc.) are discarded. During backward propagation, the complete forward computation is re-executed to reconstruct the required gradients \citep{chen2016training}. This strategy yields the maximum memory benefit, at the cost of additional forward computation, whose overhead varies across GPU types and kernel implementations. It is suitable for memory-extremely-constrained devices.
    \item \textbf{Selective Recomputation.} On modern GPUs with standard Transformer configurations, this strategy offers limited standalone benefit: its standalone memory benefit can be small, and its time overhead is difficult to distinguish from measurement noise \citep{korthikanti2023reducing}. This is primarily because optimized attention kernels and fused operations already minimize activation retention. Therefore, MemX employs it as a transient state between None and Full in hybrid pipelines.
    \item \textbf{None (No Recomputation).} All intermediate activations from forward propagation are fully preserved and directly reused during backward propagation \citep{huang2018gpipe}. Applying this strategy introduces zero computation overhead, while it has maximal memory footprint. It is suitable for memory-abundant high-performance devices, such as Huawei Ascend 910B (64GB), trading space for time to maximize compute throughput.
\end{itemize}

MemX assigns recomputation strategies at the granularity of pipeline stages. Let the cluster contain $K$ types of heterogeneous devices, where each GPU type $k$ has memory capacity $M_{k}$ and peak compute power $C_{k}$. For an $N$-layer model with $p$ PP pipeline stages, the decision variable is $r_{k,p} \in \{\text{Full}, \text{Selective}, \text{None}\}$, representing the recomputation strategy for GPU type $k$ at stage $p$. Given these decision variables, MemX selects the per-stage recomputation strategy in two levels: a hard level that guarantees every stage fits into device memory, and a soft level that picks the most cost-effective strategy among the feasible ones.

\textbf{First level: hard constraint satisfaction.} For device $k$, the feasibility check compares $\bar{D}_{k,p}^{\text{None}}$ and $D_{k,p}^{\text{Full}}$ against $M_{k}$. If $D_{k,p}^{\text{Full}} > M_{k}$, the configuration is pruned; if $\bar{D}_{k,p}^{\text{None}} \leq M_{k}$, None is retained; otherwise Full is enforced.

\textbf{Second level: joint optimization selection.} Within the feasible domain, MemX performs strategy refinement to minimize total iteration time. The key observation is that the time-space tradeoff of recomputation exhibits nonlinear characteristics and is jointly affected by device compute power and memory bandwidth. For devices with abundant compute power but limited memory bandwidth, aggressive recomputation may become bottlenecked by memory bandwidth due to frequent recomputation; for devices with scarce compute power but abundant memory bandwidth, the time penalty of aggressive recomputation is relatively smaller.

MemX establishes a marginal time-space exchange rate model for each GPU type:

\begin{equation}
\rho_{k}^{(s_{1} \to s_{2})} = \frac{\Delta T_{k}(s_{1} \to s_{2})}{\Delta M_{k}(s_{1} \to s_{2})}
\end{equation}

where $\Delta T$ denotes the time change caused by strategy switching, and $\Delta M$ denotes the memory change. When strategy levels exhibit discontinuous $\rho$ (e.g., Selective yielding near-zero benefit on some devices), MemX collapses the intermediate level and treats the effective choice as binary between None and Full.

\subsubsection{Search Algorithm Design}

\textbf{Multi-objective optimization.} MemX formulates automatic parallel search as a three-objective optimization problem. The first objective is iteration time, which directly determines training efficiency. The second is peak memory usage, defined as the maximum memory consumption across all devices in the cluster. Unlike existing systems that treat memory merely as a hard constraint, MemX explicitly incorporates it as an optimization objective to improve resource efficiency and prevent scenarios where some devices remain severely underutilized while others operate at their limits. The third is memory utilization variance, which quantifies the deviation of each device's memory usage from the cluster mean. Excessive variance indicates localized bottlenecks: memory-constrained devices are forced to adopt aggressive recomputation, significantly increasing per-layer execution time and dragging down overall pipeline throughput. These three objectives are inherently conflicting: reducing memory footprint requires aggressive recomputation at the cost of increased iteration time, while pursuing balanced utilization may sacrifice the efficiency of optimal devices.
MemX aggregates the three objectives into a single scalar objective through weighted product, which automatically penalizes short boards:

MemX aggregates the three objectives into a single scalar objective through a weighted product:

\begin{equation}
    \min_{\mathbf{x}} \quad
    \Phi(\mathbf{x}) =
    \widehat{T}(\mathbf{x})^{w_T}
    \cdot
    \widehat{M}(\mathbf{x})^{w_M}
    \cdot
    \widehat{V}(\mathbf{x})^{w_V}
\end{equation}

where
\[
\widehat{T}(\mathbf{x}) = \frac{T(\mathbf{x})}{T_{\mathrm{ref}}},
\quad
\widehat{M}(\mathbf{x}) = \frac{M(\mathbf{x})}{M_{\mathrm{ref}}},
\quad
\widehat{V}(\mathbf{x}) =
\frac{V(\mathbf{x})+\epsilon_V}{V_{\mathrm{ref}}+\epsilon_V}.
\]

Here, $T(\mathbf{x})$ denotes iteration time, $M(\mathbf{x})$ denotes peak memory usage, and $V(\mathbf{x})$ denotes the variance of memory usage across devices. The reference values $T_{\mathrm{ref}}$, $M_{\mathrm{ref}}$, and $V_{\mathrm{ref}}$ are fixed for a given candidate set, for example by using the median of all feasible candidates. This rescaling makes the three terms dimensionless without changing their ordering, because the reference denominators contribute only a constant multiplier to $\Phi$. We set $\epsilon_V$ to a small fraction of $V_{\mathrm{ref}}$ to prevent a zero-variance configuration from dominating the product solely because $V(\mathbf{x})=0$. With $w_T,w_M,w_V\geq 0$ and $w_T+w_M+w_V=1$, the weights can be interpreted as elasticities: a one-percent increase in objective dimension $i$ raises $\Phi$ by approximately $w_i$ percent.

We use three fixed weight profiles rather than tuning the weights on test data. The default heterogeneous profile $(w_T,w_M,w_V)=(0.4,0.2,0.4)$ is used for clusters with substantial device diversity, where identifying balanced configurations is as important as minimizing iteration time. The safety-first profile $(0.3,0.5,0.2)$ is used when the minimum predicted memory margin falls below 10\%, prioritizing feasibility and OOM avoidance. The throughput-first profile $(0.7,0.2,0.1)$ is used for homogeneous or weakly heterogeneous clusters, where memory pressure and imbalance are less pronounced. All profiles are selected before evaluation according to these deployment conditions. Section 4.2.4 uses the default heterogeneous profile, while Section 4.4 uses the safety-first profile because the small V100/T4 cluster has the tightest memory headroom among the evaluated systems.

\textbf{Heuristic pruning.} MemX designs three pruning rules to control the search space. First, cross-device communication restriction: tensor parallelism and data parallelism involve frequent synchronization operations that are latency-sensitive, hence their communication groups are constrained within individual pipeline stages. Second, memory overflow pruning: the memory prediction model rapidly estimates per-device memory consumption under the Full recomputation strategy, which preserves the minimal activations and constitutes the theoretical lower bound of memory usage; if this lower bound still exceeds physical capacity, the configuration is infeasible regardless of recomputation strategy and is immediately pruned, avoiding invocations of the performance prediction model on other recomputation configurations. Third, iteration time lower-bound pruning: the minimum achievable iteration time for a partial configuration is computed from device peak compute and theoretical communication volume; if this lower bound exceeds the current best solution, the branch cannot yield a superior configuration and is pruned.

\textbf{End-to-end performance.} MemX adopts the micro-batch as the minimal modeling unit, striking a balance between iteration-level accuracy and operator-level efficiency. A micro-batch encompasses complete layer computations, capturing dynamic characteristics such as operator fusion and kernel launch overheads, while maintaining controllable per-prediction overhead and aligning with the fundamental scheduling unit of pipeline parallelism. MemX supports two pipeline scheduling schemes.

\textbf{1F1B Scheduling} \citep{narayanan2019pipedream}. Let the global batch size be $B$, data parallelism degree $D_P$, and micro-batch size $b_m$. The number of micro-batches (gradient accumulation steps) is $G=B/(D_P\cdot b_m)$. Let $t_i^{mb}$ denote the single micro-batch execution time at PP stage i (including forward and backward of all layers in that stage), $t_i^{pp}$ the inter-stage PP communication time, and $t^{fix}$ the fixed overhead (e.g., optimizer step, data loading). The iteration time comprises three components: pipeline fill and drain, steady-state execution, and communication overhead. Specifically, the first micro-batch incurs fill delay through all P stages, the last micro-batch incurs drain delay for backward propagation, intermediate G-1 micro-batches progress at the bottleneck stage period in steady state, plus activation passing PP communication and DP communication.

\begin{equation}
    T_{\text{1F1B}} = (G-1) \cdot \max_{i \in [P]} t_i^{\text{mb}} + \sum_{i=0}^{P-1} t_i^{\text{mb}} + 2\sum_{i=0}^{P-1} t_i^{\text{pp}} + t^{\text{fix}} + t^{\text{dp}}
\end{equation}

where $t^{dp}=2(D_P-1)M/(D_P\cdot B_{dp})$ is the data parallelism all-reduce communication time, M the model parameter memory footprint, and $B_{dp}$ the DP communication bandwidth. The first term $(G-1)\cdot \max t_i^{mb}$ is the steady-state progression time, the second term $\sum{t_i^{mb}}$ the base computation for a single micro-batch traversing all stages, and the coefficient 2 in PP communication accounts for one forward activation pass and one backward gradient pass each.

\textbf{Interleaved 1F1B Scheduling} \citep{narayanan2021efficient}. Introduce the virtual pipeline number $N_v$ (vp\_size), where each physical stage is divided into $N_v$ virtual stages with micro-batches interleaved. A single physical stage is shared by $N_v$ virtual stages, and its single-micro-batch execution time scales ideally by $1/N_v$. Fill and drain overheads are amortized due to increased virtual stages, while micro-batches progress in $N_v$-way interleaving in the steady-state term.

\begin{equation}
\begin{aligned}
    T_{\text{Interleaved}} =&
    \left(G - 1 + \frac{1}{N_v}\right)
    \cdot
    \max_{i \in [P]} \frac{t_i^{\text{mb}}}{N_v}
    \\
    &+ \frac{2}{N_v}\sum_{i=0}^{P-1} t_i^{\text{mb}}
    + \sum_{i=0}^{P-1} t_i^{\text{pp}}
    + t^{\text{dp}}.
\end{aligned}
\end{equation}

Compared to 1F1B, the steady-state coefficient changes from $(G-1)$ to $(G-1+1/N_v)$, with the additional $1/N_v$ term reflecting extra fill from virtual stages; the base computation term is divided by $N_v$; and the bottleneck time takes $\max_i(t_i^{\text{mb}}/N_v)$. The virtual stage number $N_v$ is constrained by memory: $N_v\cdot A_{\text{stage}}\leq M_{\min}$, where $A_{\text{stage}}$ is the single-stage activation peak.

MemX supports adaptive selection among two pipeline scheduling schemes. The baseline 1F1B alternates forward and backward execution, offering simplicity and stable memory footprint at the cost of substantial pipeline bubbles. Interleaved 1F1B extends this by partitioning each physical stage into virtual stages that execute in an interleaved manner, effectively compressing bubbles while increasing activation memory overhead; this scheme is suitable for memory-abundant scenarios. The adaptive selection mechanism relies on two indices: the memory pressure index, defined as the ratio of minimum available memory to maximum stage memory demand, and the GPU heterogeneity index, quantified as the coefficient of variation of device compute power (standard deviation over mean). When memory is constrained or heterogeneity is pronounced, 1F1B is enforced to guarantee feasibility; when memory is abundant and devices are sufficiently homogeneous, Interleaved 1F1B is activated to pursue higher throughput.

\textbf{Reducing Prediction Overhead in Search.} The excessive search space leads to frequent invocations of the performance prediction model, causing search overhead to grow linearly with the number of candidate configurations. Reducing redundant evaluation has also been recognized by recent systems. AutoHAAP caches operator states to eliminate repeated latency calculations \citep{wang2026autohaap}, HARP clusters repeated modules to reduce profiling overhead \citep{liang2026harp}, AutoScout prioritizes high-impact configuration features and adaptively combines simulators with different fidelities \citep{shong2026autoscout}, and Mist reuses symbolic expressions after a single simulation pass \citep{zhu2025mist}. In comparison, MemX reduces invocation overhead from two complementary perspectives: model architecture design and prediction result reuse.

At the model architecture level, MemX employs a dual-output XGBoost predictor that produces iteration time and memory footprint in one inference call, avoiding two independent model invocations for each candidate configuration.

At the result reuse level, MemX establishes a hash-signature-based prediction cache mechanism. The signature is generated from the combined features of tensor parallel strategy, recomputation strategy and other hyperparameter for each micro-batch. New queries are first subjected to signature matching, with cache hits directly returning previously stored results. Cache invalidation conditions are strictly confined to GPU type changes (execution characteristics are non-transferable across GPUs from different vendors), model structure changes (number of layers, hidden dimensions, etc.), and batch size changes (affecting operator scheduling and memory allocation patterns). In typical search processes, many candidate configurations share identical substructures, allowing repeated predictor calls to be avoided. Section 4.3 shows that this reuse leads to near-constant search overhead when cluster-scale extensions leave the key performance parameters unchanged.

\subsection{Lightweight Performance Modeling}

This design is complementary to recent approaches that reduce modeling cost through symbolic expressions \citep{zhu2025mist}, analytical hardware models \citep{zong2025training}, profiled cost coefficients \citep{li2025malleus, li2025oases, wang2024pro}, graph-based or resource-level simulation \citep{strati2025sailor, guo2025adaptis}, or adaptive simulator ensembles \citep{shong2026autoscout}. MemX targets a different operating point: it performs sparse real measurements only for high-impact parameter combinations, augments them through validated interpolation, and trains a lightweight dual-output predictor for both time and memory.

\subsubsection{Minimal-set Profiler}
\textbf{Key Parameter Identification.} The configuration space for automatic parallel search is determined by multiple dimensions, encompassing parallel strategies, model partitioning schemes, recomputation strategies, and model hyperparameters. Exhaustively traversing all parameter combinations incurs sampling overhead that grows exponentially with dimensionality, rendering it particularly infeasible in cross-vendor heterogeneous scenarios. MemX identifies high-impact parameter subsets through ablation experiments: fixing other parameters at typical default values while varying a single parameter and observing the fluctuation amplitude of iteration time and memory footprint. Parameters with large fluctuation amplitudes are classified as high-impact and incorporated into subsequent sampling; those with small fluctuation amplitudes are classified as low-impact and only fixed typical values are selected.

The ablation experiments rely on small-scale pre-sampling: for each type of heterogeneous device, a representative server hosting that GPU type is selected to execute short-iteration measurements covering extreme and central values of each parameter. Taking LLaMA2-7B training as an example, typical high-impact parameters include TP degree (determining tensor sharding granularity and all-reduce communication volume), recomputation strategy (significantly altering the memory-time tradeoff curve), and micro-batch size (linearly scaling activation memory and affecting operator utilization); typical low-impact parameters include maximum position embedding length (only determining the upper bound of pre-allocated positional encoding tables, with actual sequence lengths during training constrained by other parameters). Through this pre-screening, MemX compresses the parameter dimensions requiring dense sampling from over ten to a core subset of 3-5 dimensions, laying the foundation for subsequent sparse sampling.

Scope of parallel dimensions: Throughout this paper, the parallelization strategies in the candidate space refer to tensor, pipeline, and data parallelism (TP/PP/DP); three other dimensions are excluded. First, sequence parallelism (SP) in Megatron-LM is not an independent device-placement dimension but a memory-saving extension of TP that shards the LayerNorm and Dropout activations along the sequence dimension\citep{shoeybi2019megatron}; it introduces no new communication groups or device mappings, and its memory effect is already absorbed by our profiled peak-memory measurements, since all profiling runs execute with SP enabled. Second, context parallelism (CP) targets extremely long sequences (tens of thousands to millions of tokens) by sharding the sequence dimension across devices, whereas our evaluated workloads use sequences of 1K--4K tokens whose activations fit in single-device memory, so CP's benefit regime does not arise. Moreover, CP is orthogonal to cross-vendor heterogeneity: it partitions within a layer and thus interacts with neither the cross-device layer placement decided by PP nor the intra-layer tensor sharding decided by TP. Third, expert parallelism (EP) applies only to Mixture-of-Experts models, while our evaluation targets dense LLMs. We note that CP and EP can be incorporated as additional device-mesh dimensions in the candidate space without structural changes to MemX's profile--predict--search pipeline, which we leave to future work.

\textbf{Sparse Sampling Strategy.} For identified high-impact parameters, MemX adopts a three-point boundary-midpoint sparse sampling strategy in lieu of exhaustive traversal. Specifically, for continuous or discrete parameters with value range $[x_{min},x_{max}]$, the minimum $x_{min}$, maximum $x_{max}$, and midpoint $(x_{min}+x_{max})/2$ or default value are sampled. The theoretical basis of this strategy lies in that most performance metrics (iteration time, memory footprint) exhibit monotonic or approximately monotonic relationships with respect to parameter variation: boundary points capture extreme behavioral characteristics, while the midpoint represents typical operating conditions, with three points sufficing to fit linear or piecewise-linear trends. For parameters with known optimal intervals (e.g., micro-batch sizes typically being powers of two), sampling points are adjusted to boundary subsets of typical values within that interval (e.g., minimum, maximum, and central values from 1, 2, 4, 8, 16, 32, 64).

Complexity reduction of the sparse sampling strategy in Algorithm 1: the reduction is substantial. For a high-impact parameter with N possible values, exhaustive sampling requires $O(N^d)$ measurements, whereas sparse sampling fixes at 3 points, reducing complexity to $O(3^d)$. Taking a 5-dimensional high-impact parameter subset as an example, if each dimension averages 10 values, exhaustive sampling requires $10^5=100000$  measurements, while sparse sampling requires merely $3^5=243$, reducing by approximately two orders of magnitude. Combined with fixed typical values for low-impact parameters, the total sampling points are compressed from a billion-scale configuration space to hundreds of key measurements, achieving order-of-magnitude reduction in profiling overhead.

\begin{figure*}
  \centering
  \includegraphics[width=0.8\textwidth]{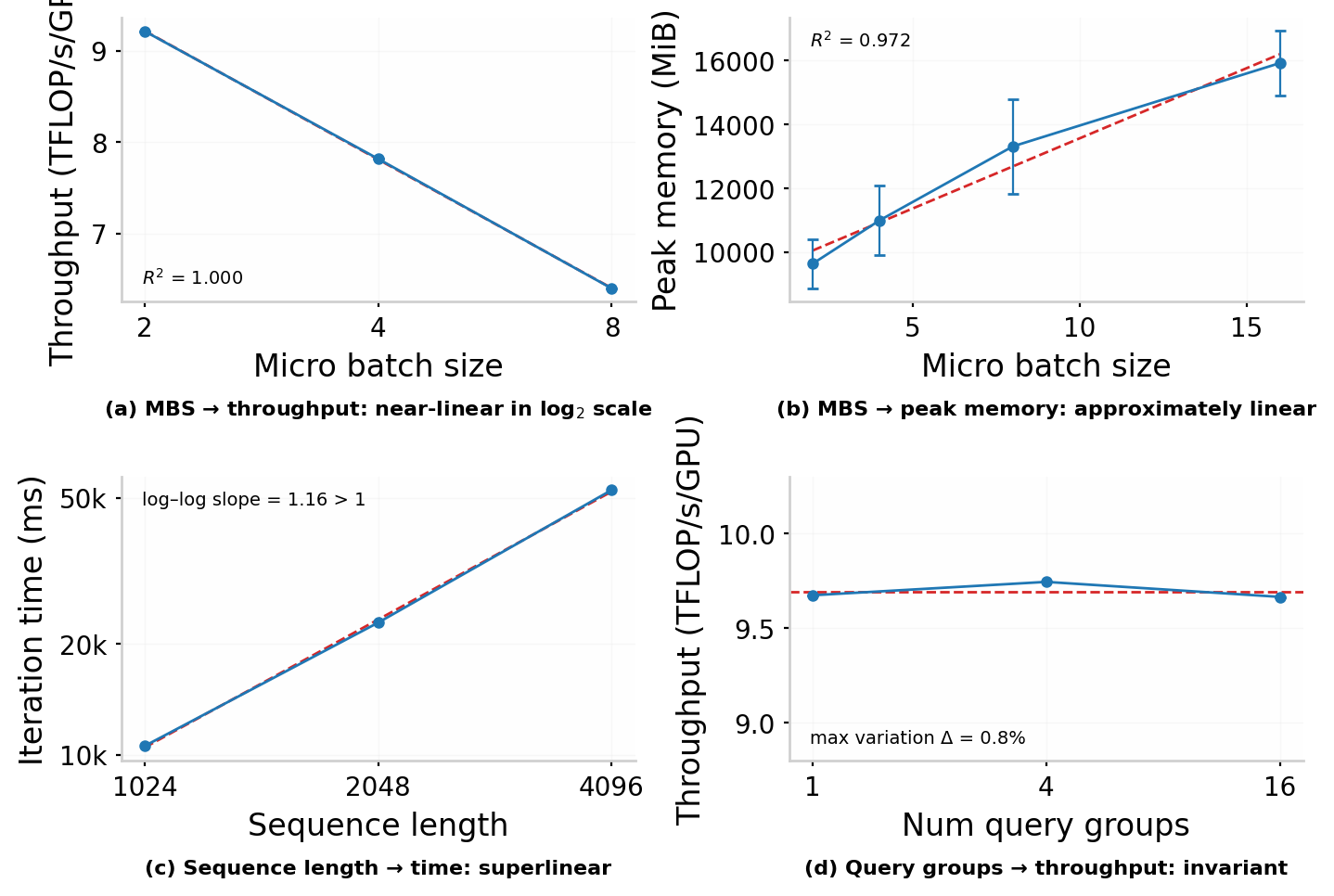}
  \caption{Measured single-parameter sweeps illustrating the trend
  classification used by the interpolation validator (Section~3.2.2),
  collected on the small two-node cluster; error bars show the standard
  deviation across ranks and repetitions, and the dashed line is a linear
  fit. (a)~Micro batch size $\rightarrow$ throughput: near-linear in
  $\log_2$ scale ($R^2=1.000$); (b)~micro batch size $\rightarrow$ peak
  memory: approximately linear ($R^2=0.972$); (c)~sequence length
  $\rightarrow$ iteration time: superlinear (log--log slope $1.16>1$),
  requiring nonlinear extrapolation; (d)~query groups $\rightarrow$
  throughput: invariant ($\Delta=0.8\%$), fixed at the default value.}
  \label{fig:interpolation-validation}
\end{figure*}

\subsubsection{Data Augmentation \& Training}
\textbf{Linear Interpolation Hypothesis.} Sparse sampling substantially reduces measurement overhead, yet the resulting sample size is insufficient to support high-precision prediction models. MemX generates augmented data based on the linear interpolation hypothesis: for certain continuous parameters (e.g., batch size, sequence length), performance metrics are assumed to vary approximately linearly between sparse sampling points, with intermediate configurations synthesized through interpolation for training.

The validity of this hypothesis is verified through small-scale experiments: on representative devices, several parameter intervals are densely sampled with 5-7 points to measure actual performance, linear models are fitted, and the coefficient of determination $R^2$ is computed. If $R^2 > 0.95$, the interval is deemed to satisfy the local linearity assumption and safe for interpolation; otherwise the interval is shrunk or piecewise-linear fitting is adopted. Experimental results in Table \ref{tab:interpolation} reveal significant variation across parameter-metric associations. "num\_layers to memory" and "micro\_batch\_size to TFLOP/s/GPU" satisfy strong linearity ($R^2 > 0.98$), enabling direct linear interpolation; "seq\_length to Iteration time" exhibits superlinear growth ($R^2 = 0.78$) due to quadratic attention complexity, requiring nonlinear extrapolation; "num\_query\_groups to all three metrics" show negligible variation ($R^2 < 0.05$), thus directly fixing default values.

\begin{table*}[width=0.75\textwidth,cols=5,pos=h]
\centering
\caption{Examples of Linear Interpolation Validation Results}
\label{tab:interpolation}
\small
\begin{tabular}{llccc}
\toprule
\textbf{Parameter} & \textbf{Metric} & \textbf{Trend} & $R^2$ & \textbf{Interpolation Strategy} \\
\midrule
num\_layers & Peak GPU memory & Strong linear & 0.998 & Large-step linear \\
micro-batch size & Peak GPU memory & Approx. linear & 0.971 & Large-step linear \\
micro-batch size & TFLOP/s/GPU & Strong linear & 1.000 & Linear \\
seq\_length & Iteration time & Superlinear & 0.780 & Nonlinear extrapolation \\
num\_query\_groups & All three metrics & Constant & $<$0.05 & Fixed default \\
\bottomrule
\end{tabular}
\end{table*}

Based on this classification, MemX dynamically adjusts how much data is generated through interpolation: strong linear cases use large steps to generate many points, near-linear cases use smaller steps or split into segments, nonlinear cases reduce the interpolation range or skip it entirely, and constant cases simply use fixed values. Figure~\ref{fig:interpolation-validation} illustrates the three trend types on measured single-parameter sweeps: near-linear parameters are interpolated with large steps, the superlinear case triggers interval shrinking, and invariant parameters are fixed without any further measurement. The augmented dataset typically grows 5-10× in size, effectively reducing overfitting risks when training samples are limited.

\textbf{XGBoost Prediction Model.} MemX employs gradient boosted trees (XGBoost) as the performance predictor, balancing prediction accuracy, inference speed, and interpretability. XGBoost iteratively adds prediction trees where each tree corrects the errors of its predecessor, and adopts second-order Taylor expansion in loss function optimization for more precise handling of non-linear problems; regularization terms are incorporated to control model complexity and prevent overfitting \citep{chen2016xgboost}. Its gradient boosting mechanism is more sensitive to feature importance, quantifying contribution through the gain of features used as split points, facilitating identification of key configuration parameters. Input features encode complete configuration information: GPU type, parallel strategy configuration, recomputation strategy, and model hyperparameters (batch size, sequence length, hidden dimension, layer count, etc.). Dual outputs are produced through a multi-task XGBoost variant: shared tree structures learn feature representations, with separate time prediction heads and memory prediction heads at leaf nodes.

Model training uses the augmented dataset, with both iteration time and memory footprint evaluated by MAPE. Hyperparameter tuning employs Bayesian optimization, efficiently exploring the space of tree depth, learning rate, and subsample ratio based on Gaussian processes \citep{snoek2012practical}. XGBoost feature importance outputs guide the search strategy: high-importance features are varied with priority, while low-importance features are fixed at defaults. The design target is to minimize MAPE with millisecond-level single prediction, satisfying the real-time demands of tens to hundreds of thousands of evaluations during search. In the evaluation of Section 4.2, MemX achieves a mean MAPE of 12.17\%.

\section{Evaluation}

\subsection{Experimental Setup}
\textbf{Hardware Environment.} We evaluate MemX on a cross-vendor heterogeneous cluster comprising 20 GPUs across three physical nodes. To study the impact of memory–performance asymmetry, the cluster is intentionally configured with GPUs from three different hardware vendors. Table \ref{tab:cluster} summarizes the hardware specifications.

\begin{table}[width=\linewidth,cols=4,pos=h]
\centering
\caption{Hardware configuration of the evaluation cluster.}
\label{tab:cluster}
\small
\begin{tabular}{lp{10mm}p{12mm}p{24mm}}
\toprule
\textbf{GPU Type} & \textbf{Quantity} & \textbf{HBM Capacity} & \textbf{Normalized Computing Power} \\
\midrule
NVIDIA A100 & 4 & 80 GB & 128 \\
Vendor-B GPU & 8 & 32 GB & 99 \\
Vendor-C GPU & 8 & 32 GB & 46 \\
\bottomrule
\end{tabular}
\end{table}

\textbf{Communication Bandwidth.} The inter-node bandwidth across all three nodes is 28 GB/s. The A100 intra-node bandwidth with NVLink is 240 GB/s. The Vendor-B and Vendor-C GPUs exhibit a notable bandwidth asymmetry: the cross-NUMA link (12 GB/s) is slower than the 4-GPU interconnect (17 GB/s), and both are substantially slower than the A100 NVLink domain. This heterogeneity in both compute and communication characteristics poses distinct challenges for load balancing and performance modeling.

\textbf{Software Environment.} We implement MemX on top of Megatron-LM 0.8 \citep{narayanan2021efficient}. All baseline methods are adapted to the same software stack to ensure a fair comparison.

\textbf{Models.} We conduct experiments on five representative LLM configurations spanning two model families and multiple scales \citep{touvron2023llama, touvron2023llama2, grattafiori2024llama}. Table \ref{tab:models} lists their architectural hyper-parameters.

\begin{table*}[width=0.7\textwidth,cols=7,pos=h]
\centering
\caption{Model configurations used in evaluation.}
\label{tab:models}
\small
\begin{tabular}{ccccccc}
\toprule
\textbf{Model} & \textbf{Layers} & \textbf{Hidden Dim} & \textbf{Heads} & \textbf{Head Dim} & \textbf{Vocab Size} & \textbf{Seq Len} \\
\midrule
LLaMA2-7B &	32 & 4096 & 32 & 128 & 32000 & 4096 \\
LLaMA2-13B & 40 & 5120 & 40 & 128 & 32000 & 4096 \\
LLaMA2-70B & 80 & 8192 & 64 & 128 & 32000 & 4096 \\
LLaMA3-8B & 32 & 4096 & 32 & 128 & 128256 & 8192 \\ 
LLaMA3-70B & 80 & 8192 & 64 & 128 & 128256 & 8192 \\
\bottomrule
\end{tabular}
\end{table*}

\textbf{Baseline Methods.} We compare MemX against four state-of-the-art automatic parallelism systems:
\begin{itemize}[]
    \item \textbf{AMP:} Searches over TP, PP, DP, and micro-batch size \citep{li2022amp}.
    \item \textbf{Metis:} Extends AMP with device-group partitioning and supports heterogeneous parallelism \citep{um2024metis}.
    \item \textbf{HexiScale:} Employs graph partitioning and analytical cost modeling \citep{yanhexiscale}.
    \item \textbf{FaSOP:} Uses a Minmax-Stage layer assignment algorithm \citep{hwang2024FaSOP}.
\end{itemize}

We focus the empirical comparison on four representative systems with directly comparable search objectives and deployment assumptions, and discuss more recent systems in Section 2. To examine behavior at larger scales, Section 4.5 further evaluates MemX on simulated 40-GPU and 60-GPU clusters derived from the measured data.

\subsection{End-to-End Comparison}

We evaluate MemX against the four baseline methods on the 20-GPU cluster described in Table \ref{tab:cluster}. The evaluation is grounded in a large real-measurement dataset collected on this cluster: 36,740 training records covering 1,628 unique configurations, spanning TP, PP, DP, micro-batch sizes, sequence lengths, and multiple recomputation and attention-grouping settings, across all three GPU types. The workload includes different models in Table \ref{tab:models}. 

For each method, we report (i) the size of the explored configuration space and the wall-clock search overhead, and (ii) the quality of its predictions and recommendations against real hardware measurements. Since FaSOP is not open-source, we re-implemented it on top of AMP's codebase with the communication-latency modeling revised according to its paper \citep{hwang2024FaSOP}; its search overhead is essentially identical to AMP's, and the AMP rows below therefore represent both methods.

\subsubsection{Search Space and Efficiency}
Table \ref{tab:e2e} summarizes the search scope and efficiency of each method. AMP, Metis, and HexiScale all restrict their search to a small subset of the configuration space to keep profiling overhead manageable. AMP explores only 14 configurations spanning TP, PP, DP, and micro-batch size (MBS). Metis expands the space to 123 configurations by incorporating device-group partitioning. HexiScale evaluates 24 configurations using its analytical cost model without profiling. In contrast, MemX searches over 2,240 configurations (160× the space of AMP) by including additional hyper-parameters such as sequence length, recomputation strategy, and attention-head configuration. After filtering out configurations predicted to exceed memory capacity (approximately 30\%), MemX retains 1628 valid configurations, still 125× the effective space of AMP.

Despite covering two orders of magnitude more configurations, MemX completes the search in 24.87 seconds, which is 62\% of the time taken by AMP. This counter-intuitive result stems from MemX's minimal-set profiler and XGBoost-based predictor, which replace expensive per-configuration profiling with millisecond-level inference. FaSOP does not perform an explicit search in the same sense; instead, it relies on an end-to-end cost model with greedy layer assignment, so we use the same scheme as AMP for FaSOP and omit its search time in this comparison.

\begin{table*}[width=0.82\textwidth,cols=5,pos=h]
\centering
\caption{Search space and efficiency comparison (LLaMA2/3, 20 GPUs).}
\label{tab:e2e}
\small
\begin{tabular}{ccccc}
\toprule
\textbf{Method} & \textbf{Configs Explored} & \textbf{Valid Configs} & \textbf{Search Time (s)} & \textbf{Avg. Time per Config (ms)}\\
\midrule
AMP (FaSOP) & 14 & 13 & 40.07 & 3,082 \\
Metis & 123 & 8 & 2.43 & 304 \\
HexiScale & 24 & 18 & 0.253 & 14 \\
MemX & 2,240 & 1,628 & 24.87 & 15.3 \\
\bottomrule
\end{tabular}
\end{table*}

\subsubsection{Prediction Accuracy}

Table \ref{tab:accuracy} reports the prediction accuracy of each method against real iteration times. We measure the Mean Absolute Percentage Error (MAPE) and the proportion of predictions within acceptable error bounds (Good: <20\%; Severe: >50\%).

\begin{table}[width=\linewidth,cols=5,pos=h]
\centering
\caption{Prediction accuracy comparison against real measurements.}
\label{tab:accuracy}
\small
\begin{tabular}{cccc}
\toprule
\textbf{Method} & \textbf{MAPE} & \textbf{Good (<20\%)} & \textbf{Severe (>50\%)}\\
\midrule
AMP & 66.44\% & 23.1\% & 61.5\% \\
Metis & 14.91\% & 62.5\% & 0\% \\
HexiScale & 43.52\% & 29.2\% & 50\% \\
FaSOP & 54.40\% & 30.8\% & 61.5\% \\
MemX & 12.17\% & 69.1\% & 0\% \\
\bottomrule
\end{tabular}
\end{table}

MemX achieves the best prediction accuracy across the board. Its mean MAPE of
12.17\% improves upon the next-best method Metis by 18.4\% in
relative terms, i.e., $(14.91-12.17)/14.91$, which is the improvement claimed
in our contributions; its maximum error is only 29.77\%. MemX is also one of the methods with no configuration exceeding 50\% error. In contrast, half of HexiScale's configurations suffer errors above 50\%, and 61.5\% of FaSOP's configurations fall within the >50\% error band; on its worst configuration (TP4/PP5/DP1, MBS=4), FaSOP predicts an iteration time more than twice the real measurement (114.6\% error).

In addition, MemX's accuracy advantage is structural rather than incidental. The baselines' errors degrade markedly along configuration dimensions such as micro-batch size, pipeline depth, and sequence length—for example, FaSOP's MAPE rises to 76.47\% on the MBS=2 subset and 62.07\% on the seq\_length=4096 subset. MemX's error distribution, by contrast, remains the flattest across all configuration subspaces, with its per-dimension mean MAPE never exceeding 18\%, indicating that its performance model is robust to variations in parallel structure and workload.

\subsubsection{Ranking Consistency and Practical Utility}
The end product of configuration search is a ranked recommendation list, so ranking quality is closer to practical utility than absolute accuracy. Table \ref{tab:ranktime} reports the consistency between each method's predicted ranking (by predicted iteration time) and the real ranking (by measured iteration time): Spearman's $\rho$, Kendall's $\tau$, and Top-k hit rates.

\begin{table*}[width=0.6\textwidth,cols=7,pos=h]
\centering
\caption{Ranking consistency against real time-based ranking.}
\label{tab:ranktime}
\begin{tabular}{ccccccc}
\toprule
Method & Spearman $\rho$ & Kendall $\tau$& Top-1 & Top-3 & Top-5 & Top-10 \\
\midrule
AMP       & 0.3132 & 0.1795& 0\% & 0\%   & 60\% & --- \\
Metis     & 0.8748 & 0.6939 & 0\% & 66.7\% & --- & --- \\
HexiScale & 0.4967 & 0.3835 & 0\% & 33.3\% & 20\% & 40\% \\
FaSOP     & 0.4960 & 0.3726 & 0\% & 0\%    & 20\% & ---\\
\textbf{MemX} & \textbf{0.8964} & \textbf{0.7335} & \textbf{100\%} & 66.7\% & \textbf{60\%} & \textbf{70\%}\\
\bottomrule
\end{tabular}
\end{table*}

MemX attains the strongest ranking consistency, slightly ahead of Metis and well ahead of HexiScale, FaSOP and AMP. The decisive difference, however, lies in recommendation quality: MemX is the only method whose top recommendation is the real optimum, whereas all four baselines miss it; even Metis, whose overall correlation is strong, does not hit it either. For a configuration search system, a missed Top-1 means the entire search cost yields a suboptimal configuration, so this gap is of direct practical consequence.

\subsubsection{Ranking Consistency under Memory-Balance Objective
}
We next examine whether each method's ranking can identify configurations with balanced memory usage. We re-rank the real configurations by their measured $\Phi(\mathbf{x})$ under the default heterogeneous weight profile $(w_T,w_M,w_V)=(0.4,0.2,0.4)$ defined in Section 3.1.2. This profile is selected a priori and is not tuned on the measured rankings. Results are shown in Table \ref{tab:rankphi}.

\begin{table*}[width=0.6\textwidth,cols=7,pos=h]
\centering
\caption{Ranking consistency against real memory-balance ($\Phi$) ranking.}
\label{tab:rankphi}
\begin{tabular}{ccccccc}
\toprule
Method & Spearman $\rho$ & Kendall $\tau$ & Top-1 & Top-3 & Top-5 & Top-10 \\
\midrule
AMP       & 0.1264  & 0.0256  & 0\% & 0\% & 0\% & 40\% \\
Metis     & -0.0692 & -0.0292 & 0\% & 0\% & 0\%  & ---  \\
HexiScale & 0.0372  & 0.0381  & 0\% & 0\% & 0\%  & 20\%  \\
FaSOP     & -0.0025 & 0.0008  & 0\% & 0\% & 0\%  & 40\%  \\
\textbf{MemX} & \textbf{0.9285} & \textbf{0.7797} & \textbf{100\%} & \textbf{66.7\%} & \textbf{80\%} & \textbf{80\%}\\
\bottomrule
\end{tabular}
\end{table*}

\begin{figure}
  \centering
  \includegraphics[width=0.8\linewidth]{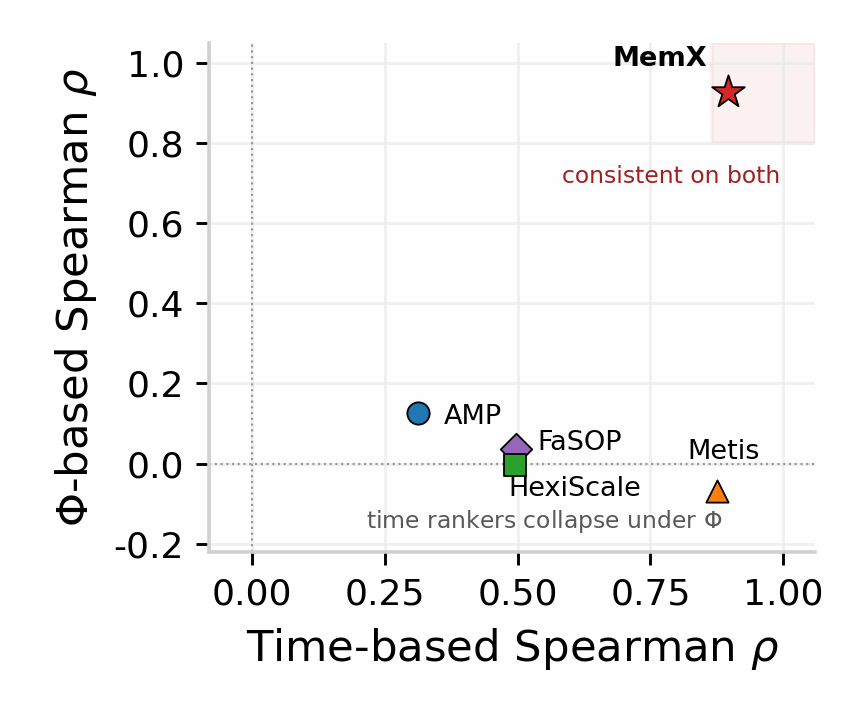}
  \caption{Time-based versus $\Phi$-based ranking consistency (Spearman's $\rho$,
  from Tables~7 and~8). The baselines' time-ranking ability varies widely
  ($0.31$--$0.87$), yet all of them collapse to near-zero correlation under
  the memory-balance objective $\Phi$; MemX is the only method that remains
  consistent on both axes.}
  \label{fig:ranking-decoupling}
\end{figure}

All baselines degenerate to random under the memory-balance objective. Their Spearman correlations fall within [-0.069, 0.126] and are all statistically insignificant. Moreover, this failure stems from a missing objective, not from poor prediction. Comparing Tables \ref{tab:ranktime} and \ref{tab:rankphi}, the baselines' time-based ranking ability varies widely from 0.31--0.87, yet their $\Phi$-based ranking ability uniformly vanishes, and the two are completely decoupled. Figure~\ref{fig:ranking-decoupling} visualizes this decoupling: time-based and $\Phi$-based ranking ability are orthogonal across all evaluated methods, and only MemX occupies the consistent-on-both region. For example, the strongest time ranker, Metis ($\rho=0.8748$), scores only $\rho=-0.0692$ under $\Phi$. In other words, even a cost model that predicts iteration time well cannot identify memory-balanced configurations.

\subsection{Scalability of Search Overhead}

\begin{figure}
\centering
\includegraphics[width=0.7\linewidth]{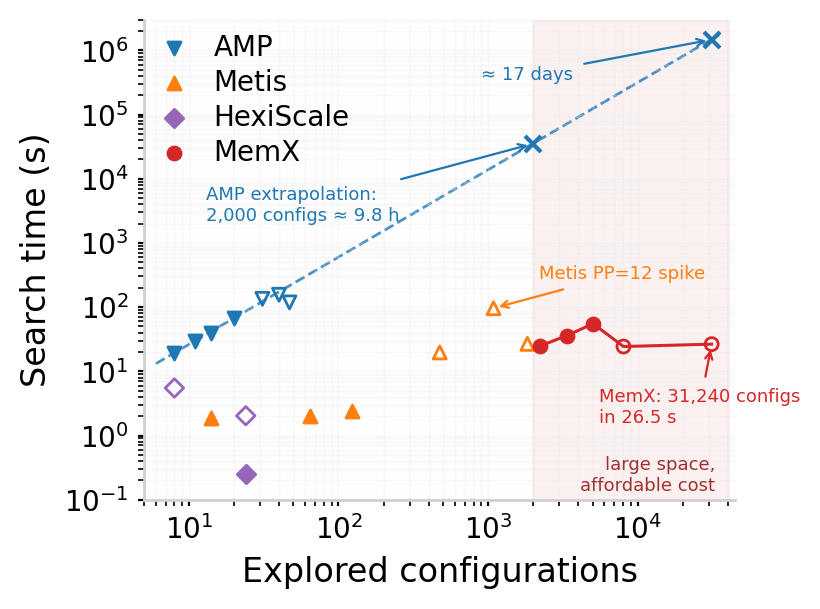}
\caption{Search time versus search-space size (log--log), combining all extension measurements of Table \ref{tab:scalability}. Filled and open markers denote extensions along the batch dimension and the cluster scale, respectively. The dashed curve is a power-law fit $t = 1.147 n^{1.36}$ ($R^2 = 0.997$) of AMP's measurements, extrapolated to 2,000 configurations ($\sim$9.8 hours). The shaded region marks large spaces ($\geq$ 2{,}000 configurations) at affordable cost---only MemX operates there, searching 31,240 configurations in 26.5 s.}
\label{fig:scalability}
\end{figure}

A practical configuration search method must remain affordable as the search space grows. In this subsection, we extend each method's search space along two axes and measure the resulting search wall-clock time: (i) the hyperparameter dimension (e.g., enlarging the global/micro-batch size options) and (ii) the cluster scale (increasing the number of nodes, which enlarges the feasible parallel-strategy space, most notably PP). Table \ref{tab:scalability} summarizes the results and Figure \ref{fig:scalability} visualizes the scaling trends.

\begin{table}[width=\linewidth,cols=4,pos=h]
\centering
\caption{Search overhead under search-space extension.}
\label{tab:scalability}
\begin{tabular}{p{12mm}llp{11mm}}
\toprule
Method & Extended setting & Configs & Search time (s) \\
\midrule
AMP / FaSOP & base (Table \ref{tab:e2e}) & 14  & 40.07  \\
    & $--$ GBS            & 8   & 19.64  \\
    & $-$ GBS            & 11  & 29.78  \\
    & + GBS           & 20  & 68.69  \\
    & + GPUs           & 47  & 118.4  \\
    & ++ GPUs          & 31  & 134.7  \\
    & +++ GPUs          & 40  & 155.6  \\
\midrule
Metis & base              & 123 & 2.43   \\
    & $---$ TP/MBS options    & 0$^{*}$ & 1.82 \\
    & $--$ TP/MBS options    & 14   & 1.88   \\
    & $-$ TP/MBS options    & 64   & 2.02   \\
    & + PP size              & 475  & 19.69  \\
    & ++ PP size             & 1,087 & 96.18 \\
    & +++ PP size             & 1,831 & 26.76 \\
\midrule
HexiScale & base & 24  & 0.253  \\
    & + nodes              & 24  & 2.03   \\
    & ++ nodes             & 8   & 5.5    \\
\midrule
\textbf{MemX} & base      & 2,240  & 24.87 \\
    & + MBS options       & 3,360  & 36.11 \\
    & ++ MBS options      & 5,040  & 55.13 \\
    & + parallel configs  & 8,040  & 24.34 \\
    & ++ parallel configs & 31,240 & 26.5  \\
\bottomrule
\multicolumn{4}{l}{\footnotesize $^{*}$ This extension yields no valid configuration}
\end{tabular}
\end{table}

The baselines' search cost grows rapidly, or unpredictably, with the space size. Because AMP (and FaSOP) profiles every configuration end-to-end, its cost grows as a power law along the batch dimension: a power-law fit $t=1.147n^{1.36}$ ($R^2=0.997$) of its measurements extrapolates to approximately 35,150 s (9.8 hours) for merely 2,000 configurations. Metis stays nearly flat when adding TP/MBS options but degrades unpredictably with PP: extending PP alters its layer-assignment strategy, which can generate large numbers of invalid configurations that must be evaluated and filtered out, spiking the search time to 96.18 s at PP=12. HexiScale's analytical model keeps parameter additions cheap, yet its overhead rises with cluster scale (0.253 s to 5.5 s from 20 to 80 GPUs) while its explored space remains the narrowest (as few as 8 valid configurations).

\begin{table*}[width=\textwidth,cols=7,pos=h]
\centering
\caption{Deployment validation on the small cluster. Baselines recommend only
the parallel dimensions and are deployed with recompute\,=\,none; their entries
span the measured variants of the remaining dimensions. The best deployed
variant of each method is shown in the time columns.}
\label{tab:deployment}
\begin{tabular}{lp{50mm}rrrrc}
\toprule
Method & Recommended configuration & Time (ms) & Time rank & $\Phi$ rank & Mem.\ var.\ (MiB$^2$) & OOM \\
\midrule
AMP       & TP2/DP1/PP3, MBS=2 & 9,211--9,912  & 1--19  & 12--26   & $5.96\times10^{5}$--$9.34\times10^{5}$  & No  \\
FaSOP     & (same as AMP)      & (same as AMP) & (same) & (same)   & (same)          & No  \\
Metis     & TP2/DP1/PP3, MBS=4 & 10,459--11,685 & 28--60 & 67--138  & $1.37\times10^{6}$--$2.91\times10^{6}$  & No  \\
HexiScale & TP1/DP2/PP3, MBS=2 & ---           & ---    & ---      & ---             & Yes \\
\textbf{MemX} & TP2/DP1/PP3, MBS=2, full recomputation, query groups=16, ... & 10,066 & 26 & \textbf{1} & $4.19\times10^{5}$ & No \\
\bottomrule
\end{tabular}
\end{table*}

\begin{figure}
  \centering
  \includegraphics[width=0.8\linewidth]{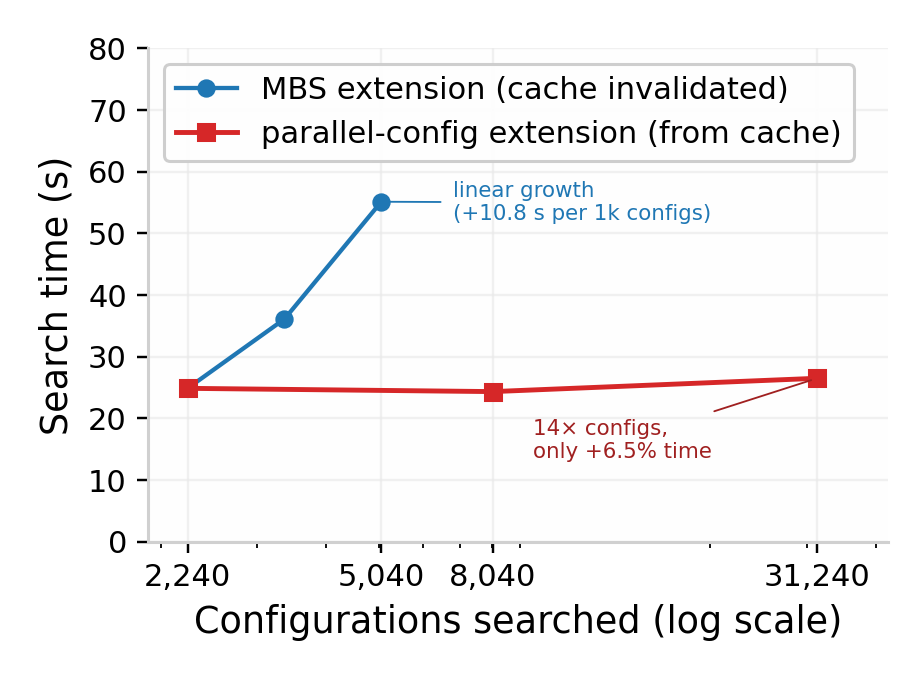}
  \caption{MemX's search time when the search space is extended along two axes (Table~9). Adding MBS options changes a key performance parameter, invalidating the hash-signature cache, so the search time grows linearly with the space size; extending only the parallel-strategy dimensions (TP/PP/DP) leaves the key parameters intact, so the additional configurations are served almost entirely from cached predictions---14$\times$ more configurations for only 6.5\% more time.}
  \label{fig:cache-effect}
\end{figure}

In contrast, MemX's search overhead is decoupled from the combinatorial growth of the parallel-strategy space, thanks to the caching mechanism described in Section 3.1.2. When the key performance parameters such as MBS of a configuration are unchanged, MemX reuses the cached prediction instead of re-invoking the XGBoost model. Extending the cluster scale only adds PP/DP combinations while leaving these key parameters intact, so the additional configurations are evaluated almost entirely from cache: 31,240 configurations, about 14$\times$ of the base space, are searched in 26.5 s, only 6.5\% more time than the base space. Along the batch dimension, however, adding MBS options changes a key performance parameter and invalidates the cache, requiring new predictor calls for the affected configurations; the cost therefore grows linearly. Figure~\ref{fig:cache-effect} contrasts the two extension axes directly: the same predictor and cache make cluster-scale extension nearly free, while batch-dimension extension is the only axis that costs new predictor calls. MemX thus searches spaces orders of magnitude larger at a fraction of the cost: at 2,000 configurations it is over 1,400$\times$ faster than AMP/FaSOP's extrapolated time, and at 31,240 configurations the contrast reaches roughly 17 days versus 26.5 s. This affordability is what enables the broad coverage and the accuracy and ranking advantages reported in Section 4.2.

\begin{figure}
\centering
\includegraphics[width=\linewidth]{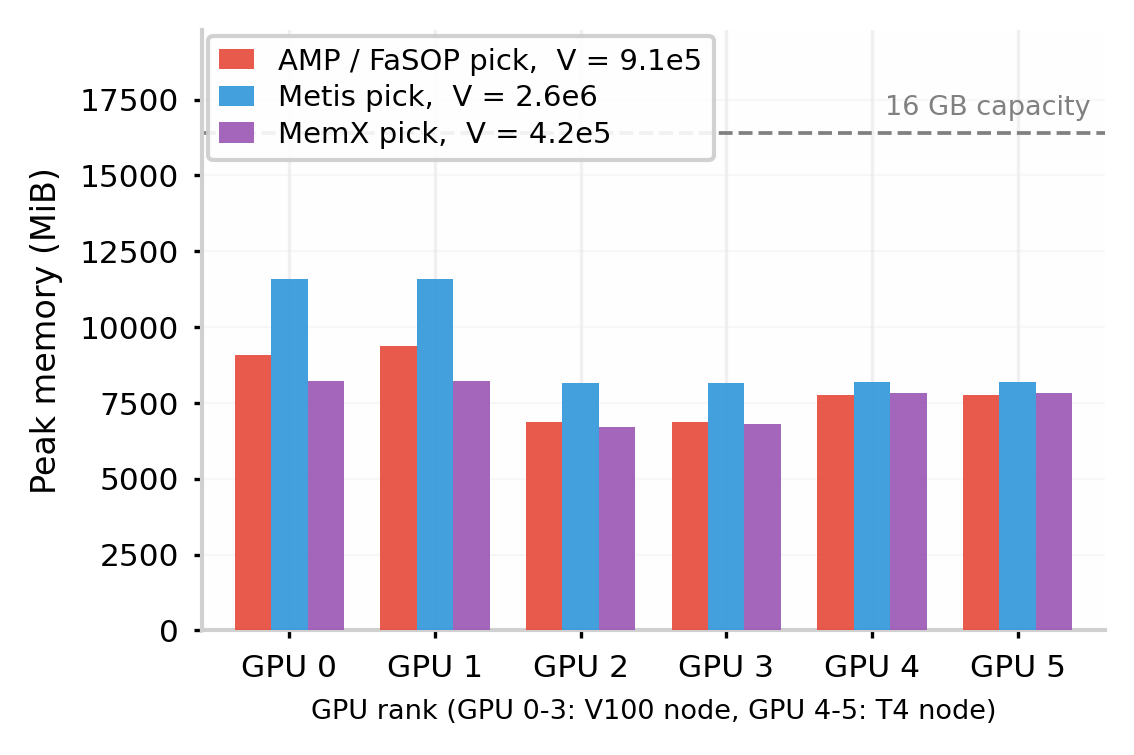}
\caption{Per-GPU peak memory of the deployed recommendations on the small cluster. MemX's pick (full recomputation) keeps every GPU at a low and even occupancy (variance $4.2\times10^{5}$), while the deployed AMP/FaSOP and Metis picks (recompute\,=\,none) are both higher and more imbalanced (variance $9.1\times10^{5}$ and $2.6\times10^{6}$, respectively).}
\label{fig:deployment_memory}
\end{figure}

\subsection{Deployment Validation on Small Cluster}

The end-to-end comparison in Section 4.2 evaluates prediction and ranking quality as proxy metrics. To verify that these metrics translate into real gains, we conduct a deployment validation on a small two-node heterogeneous cluster (one 4$\times$V100 node and one 2$\times$T4 node, 6 GPUs in total), training a 6-layer model based on LLaMA-2. The configuration space covers PP, TP, DP, MBS, sequence lengths, four recomputation strategies (none, selective, full-block, full-uniform), and several attention-head and query-group settings. We collected 4,002 per-rank measurements from 667 runs, covering 226 unique configurations with about 3 repetitions each. Each method recommends its best configuration; since some baselines recommend only the parallel dimensions (TP, PP, DP, MBS), we deploy their recommendations with recompute = none and report the resulting range over the measured variants of the remaining dimensions. The memory-balance objective $\Phi$ is computed with the safety-first weight setting $(w^T=0.3, w^M=0.5, w^V=0.2)$. The results are summarized in Table \ref{tab:deployment}.

\begin{figure}
  \centering
  \includegraphics[width=0.8\linewidth]{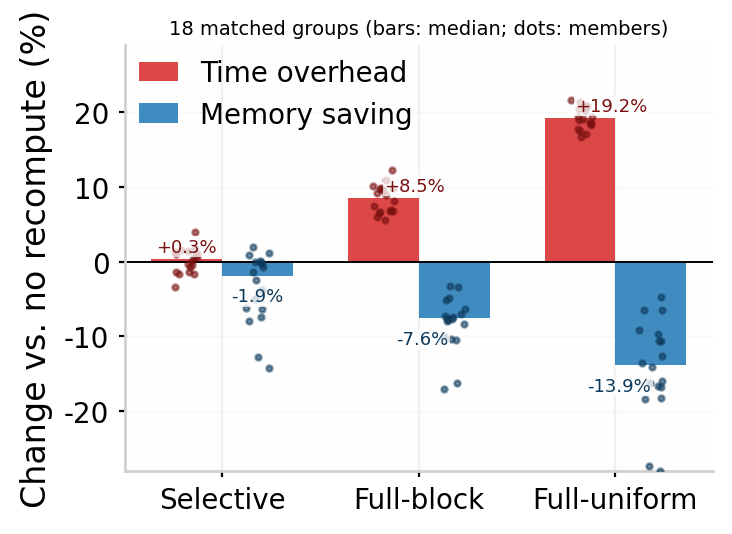}
  \caption{Time overhead versus peak-memory saving of each recomputation
  strategy relative to recompute\,=\,none, over 18 matched configuration
  groups on the small cluster (same parallel skeleton and hyperparameters;
  only the recomputation strategy varies). Full-block recomputation
  (MemX's deployed choice) saves 7.6\% peak memory for a median 8.5\% time
  overhead, a milder trade than full-uniform (13.9\% saving for 19.2\%),
  while selective recomputation is nearly free but saves little.}
  \label{fig:recompute-tradeoff}
\end{figure}

Under the iteration-time objective, the deployment result is a tie: AMP, FaSOP, and MemX all identify the true fastest configuration (9,211 ms), showing that the baselines do not fall behind on a simple objective. Metis recommends a slower MBS=4 variant (only 28th to 60th fastest), and HexiScale's recommendation cannot even be run as specified: without recomputation, a dimension HexiScale does not search, its configuration exhausts memory and OOMs. However, the time-optimal configuration itself ranks only 20th under $\Phi$, with a memory variance of $9.13\times10^{5}$, more than twice that of MemX's pick. Figure~\ref{fig:recompute-tradeoff} quantifies this trade on 18 matched configuration groups: recomputation buys memory balance at a bounded, predictable time cost, and full-block recomputation sits on the mild end of that trade, which is exactly the point MemX's $\Phi$ objective exploits.

\begin{figure}
  \centering
  \includegraphics[width=\linewidth]{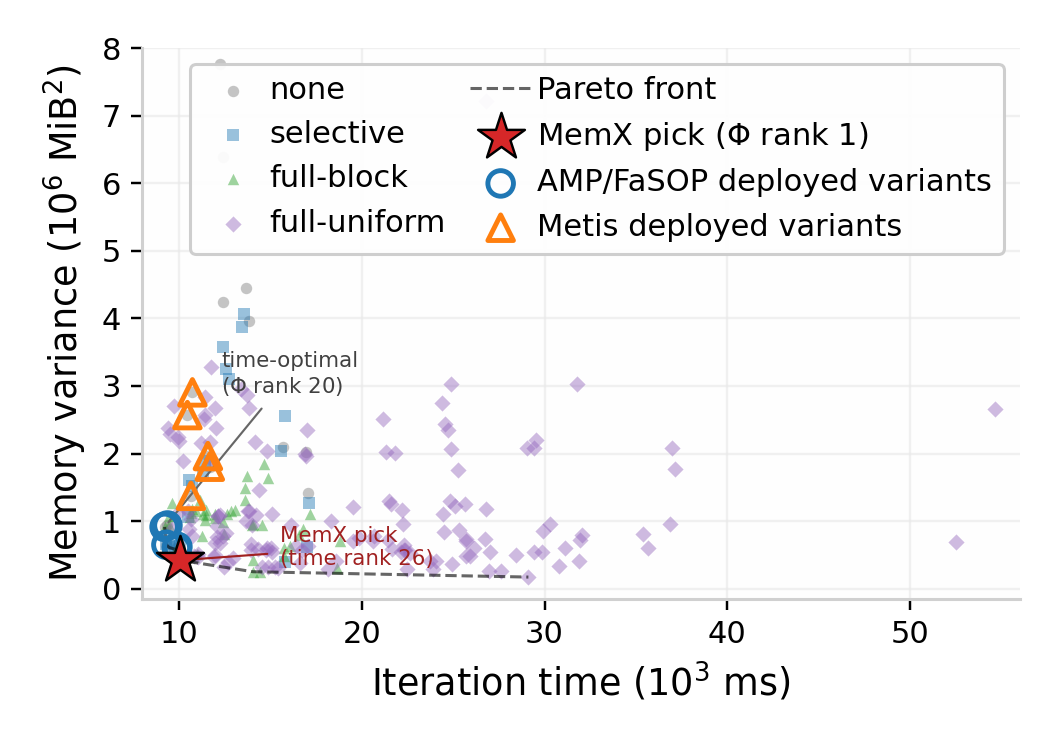}
  \caption{Iteration time versus cross-GPU memory variance for all 226 measured
  configurations on the small cluster, colored by recomputation strategy; the
  dashed line is the empirical Pareto front. MemX's deployed pick (star) attains
  the lowest memory variance ($\Phi$ rank~1) at time rank~26, only 9.3\% slower
  than the time-optimal configuration, which itself ranks only 20th under
  $\Phi$. The deployed AMP/FaSOP and Metis variants (recompute\,=\,none) cluster
  at higher variance.}
  \label{fig:sc-pareto}
\end{figure}

Under the memory-balance objective, only MemX's recommendation matches the real optimum. Figure~\ref{fig:deployment_memory} represents per-GPU peak memory of the deployed recommendations on the small cluster. Its pick shares the exact same parallel skeleton as AMP and FaSOP (TP2/DP1/PP3, MBS=2), but adds full-block recomputation, sequence length 1024, and 64 attention heads with 16 query groups, none of which the baselines can express. At 10,066 ms it is only 9.3\% slower than the absolute fastest configuration, yet it attains the lowest and most even memory footprint with mean peak 7,587 MiB and variance $4.19\times10^{5}$, ranking 1st in $\Phi$ out of all 226 configurations. AMP and FaSOP, blind to these dimensions, land at $\Phi$ ranks 12-26 depending on the deployed variant; Metis's pick lands at 67-138 with up to 7x the memory variance of MemX's pick; and HexiScale's pick does not run at all. Across all 226 configurations, the correlation between the time ranking and the $\Phi$ ranking is only 0.6568, confirming on this cluster as well that the two objectives cannot proxy each other. Figure~\ref{fig:sc-pareto} maps the full measured space: fast configurations are scattered across a wide range of memory variance, so a time-only search has no signal to land on the balanced region that MemX targets. 

In summary, MemX is the only method that wins under both criteria: tied for Top-1 on time, and the sole Top-1 on $\Phi$ with the most balanced memory usage.

\begin{table*}[width=0.66\textwidth]
\centering
\caption{Prediction accuracy and ranking consistency of MemX on simulated larger clusters extended from the measured 20-GPU data.}
\label{tab:simulated-scaling}
\begin{tabular}{lcccc}
\toprule
Cluster & Time MAPE & Memory MAPE & Spearman $\rho$ & Kendall $\tau$ \\
\midrule
20 GPUs (measured)  & 12.17\% & 6.44\% & 0.8964 & 0.7335 \\
40 GPUs (simulated) & 14.22\% & 6.65\% & 0.8876 & 0.7123 \\
60 GPUs (simulated) & 16.19\% & 6.56\% & 0.8693 & 0.6830 \\
\bottomrule
\end{tabular}
\end{table*}

\subsection{Simulated Scaling to Larger Clusters}

The previous experiments are conducted on a 20-GPU cross-vendor cluster and a 6-GPU small cluster. Since physical cross-vendor resources are limited, we further prove MemX's prediction and ranking quality hold at larger scales of clusters through a simulation experiment by extending the measured 20-GPU cluster data to 40-GPU and 60-GPU clusters.

The extension proceeds in two steps. First, configuration extension: each measured configuration is extended along the pipeline dimension only---a base configuration with parallel strategy (TP, PP, DP) is mapped to extended configurations with PP in {2p, 3p} (e.g., PP = 5 extends to 10 or 15 stages, and PP = 10 to 20 or 30 stages, covering 40-rank and 60-rank layouts), while TP, DP, and all other hyper-parameters remain unchanged. The LLM model depth grows proportionally with PP, so that the per-stage layer count is preserved while the per-stage execution time and memory footprint of such an extended configuration are inherited from the corresponding measured stages. Second, cluster extension: the added GPUs follow the original vendor mix (approximately 1:2:2 across the three GPU types, i.e., 8/16/16 for 40 GPUs and 12/24/24 for 60 GPUs). Within each extension scheme, the mapping between pipeline stages, GPU types, and ranks is fixed, and tensor-parallel groups are always formed within a single GPU type (no heterogeneous TP). Stage memory usage is taken directly from the measured values, and stage execution time is scaled by the normalized compute-power ratio of the target GPU type (Table 3) relative to the measured one. This procedure adds 10,036 derived configurations to the evaluation set.

\begin{figure}
  \centering
  \includegraphics[width=\linewidth]{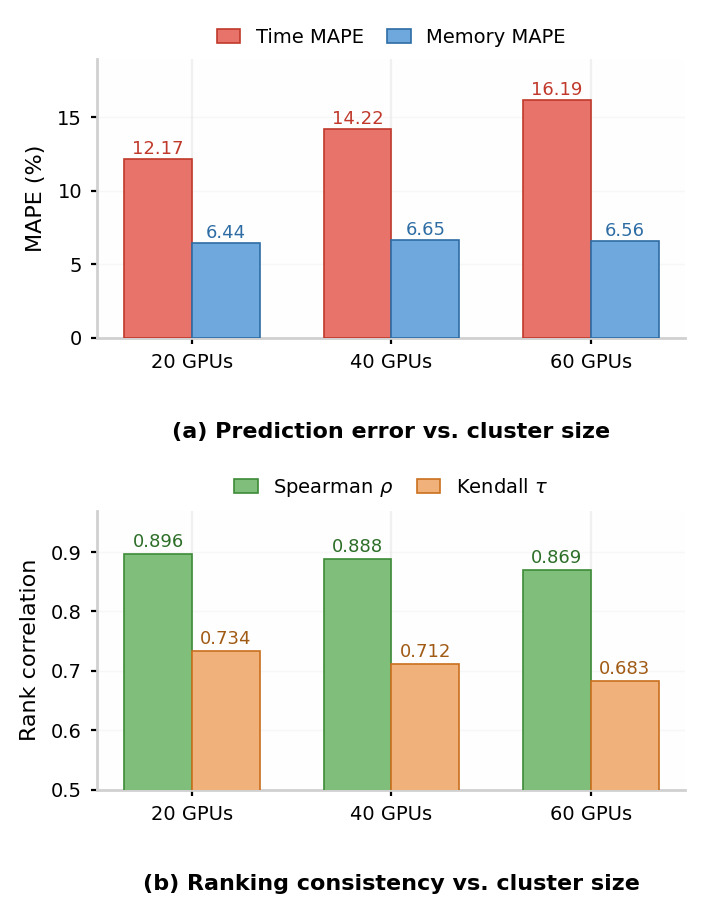}
  \caption{Prediction accuracy and ranking consistency of MemX under
  simulated scaling from 20 to 60 GPUs (Section~4.5). (a)~Mean MAPE grows
  mildly from 12.17\% to 16.19\% while the median stays flat;
  (b)~Spearman's $\rho \geq 0.86$ and Kendall's $\tau \geq 0.68$ throughout,
  indicating that ranking quality degrades gracefully with cluster size.}
  \label{fig:simulated-scaling}
\end{figure}

Table \ref{tab:simulated-scaling} and Figure \ref{fig:simulated-scaling} report the results of Memx's simulation. Prediction accuracy degrades gracefully: time MAPE grows from 12.17\% at 20 GPUs to 14.22\% at 40 GPUs and 14.91\% at 60 GPUs, while memory MAPE remains stable around 6.5\%. Ranking quality degrades much more slowly: Spearman's rho stays at or above 0.87 and Kendall's tau at or above 0.69 at all scales. The mild degradation is expected, because PP extension lengthens the pipeline and accumulates per-stage prediction error along the critical path, whereas TP/DP dimensions (which dominate the remaining configuration diversity) are inherited without rescaling. These results indicate that MemX's modeling pipeline remains consistent under PP-dominant scaling, complementing the search-overhead scalability shown in Section 4.3.

Although Time MAPE increases as the simulation scale expands, we argue that the error will not continue to grow linearly to unacceptable levels at extreme scales (e.g., 1000$+$ GPUs). First, error propagation along the PP critical path is sub-linear: per-stage errors are largely uncorrelated and partially cancel out, so the relative error grows slower than pipeline depth. Second, scaling to 1000 GPUs in practice increases both TP and DP proportionally, which introduces comparatively smaller error than PP; thus, our PP-dominant scaling experiments represent a conservative upper bound. Most importantly, configuration search depends on ranking quality rather than absolute accuracy. Even under the worst-case scenario in our current experiments, Spearman's $\rho$ remains above 0.85, ensuring the utility of search results.

Besides, we note that the extended ground truth is derived from measured stages under compute-ratio scaling. The study therefore validates the internal consistency of MemX's modeling pipeline rather than replacing real large-cluster validation; effects absent from the derivation, such as emerging cross-node communication patterns at 60 GPUs, are not captured.

\section{Conclusion and Future Work}
We propose MemX, a memory-aware automatic parallelism system for cross-vendor heterogeneous GPU training. MemX formulates configuration search as a joint optimization problem over iteration time, peak memory usage, and cross-device memory variance, and supports heterogeneous recomputation strategies at pipeline-stage granularity. To make the search cost affordable, MemX combines a minimal-set profiler, interpolation-based data augmentation, a dual-output XGBoost predictor, and a hash-signature prediction cache. Experiments against AMP, Metis, HexiScale, and FaSOP show that MemX achieves the best prediction accuracy and ranking consistency, and is the only method that identifies the real Top-1 configuration under both time-based and memory-balance objectives. It also searches substantially larger configuration spaces at much lower cost. Deployment on a small heterogeneous cluster confirms that the predicted memory-balance advantage of MemX finds a runnable, more balanced configuration.

The MemX evaluation in this paper focuses on LLaMA-family models and two moderate-scale clusters, and the objective weights are selected from preset deployment profiles rather than learned online. Future work includes evaluating MemX on larger production clusters with 1000+ GPUs and more model families, studying adaptive weight selection under changing workload and memory conditions, and integrating the memory-aware configurations with cluster-level scheduling and multi-tenant resource management.

Beyond those incremental extensions, a longer-term direction is to automate the entire automatic-parallelism workflow rather than individual decisions within it. A multi-agent scheme could assign profiling, performance modeling, configuration search, and deployment validation to cooperating agents that exchange intermediate results and iteratively refine the configuration search process with little human intervention. Such automation, however, introduces its own systems challenges. Multi-agent orchestration adds non-trivial overhead: inter-agent communication, repeated model invocation, and reasoning latency can easily exceed the wall-clock cost of the deterministic search itself, which in MemX completes within tens of seconds. Reliability is a further concern that LLM-driven agents may produce inconsistent or hard-to-reproduce choices, as an erroneous intermediate decision by a single agent can propagate through the pipeline. A practical design would therefore keep fast, deterministic components as the execution substrate (e.g., MemX's predictor and prediction cache), and delegate only high-level planning, bottleneck diagnosis, and constraint negotiation to agents, where their flexibility outweighs their cost.

\section*{Acknowledgments}

This work was supported by 2023YFB4503904 of MOST in China.

\printcredits
\bibliographystyle{cas-model2-names}

\bibliography{cas-refs-r1}

@misc{MI300X,
  author = {{Advanced Micro Devices, Inc.}},
  title = {{AMD Instinct™ MI300X Accelerators}},
  howpublished = {\url{https://www.amd.com/en/products/accelerators/instinct/mi300/mi300x.html}},
}

@misc{Gaudi,
  author = {{Intel Corporation}},
  title = {{Intel Gaudi 3 AI Accelerators}},
  howpublished = {\url{https://www.intel.com/content/www/us/en/products/details/processors/ai-accelerators/gaudi.html}},
}

@misc{biv150,
  author = {{Iluvatar CoreX}},
  title = {{Iluvatar BI-V150}},
  howpublished = {\url{https://www.iluvatar.com/productDetails?fullCode=cpjs-yj-xlxl-tg150}},
}

@article{singh2025openai,
  title={Openai gpt-5 system card},
  author={Singh, Aaditya and Fry, Adam and Perelman, Adam and Tart, Adam and Ganesh, Adi and El-Kishky, Ahmed and McLaughlin, Aidan and Low, Aiden and Ostrow, AJ and Ananthram, Akhila and others},
  journal={arXiv preprint arXiv:2601.03267},
  year={2025}
}

@article{zeng2026glm,
  title={Glm-5: from vibe coding to agentic engineering},
  author={Zeng, Aohan and Lv, Xin and Hou, Zhenyu and Du, Zhengxiao and Zheng, Qinkai and Chen, Bin and Yin, Da and Ge, Chendi and Huang, Chenghua and Xie, Chengxing and others},
  journal={arXiv preprint arXiv:2602.15763},
  year={2026}
}

@article{grattafiori2024llama,
  title={The llama 3 herd of models},
  author={Grattafiori, Aaron and Dubey, Abhimanyu and Jauhri, Abhinav and Pandey, Abhinav and Kadian, Abhishek and Al-Dahle, Ahmad and Letman, Aiesha and Mathur, Akhil and Schelten, Alan and Vaughan, Alex and others},
  journal={arXiv preprint arXiv:2407.21783},
  year={2024}
}

@article{li2022amp,
  title={Amp: Automatically finding model parallel strategies with heterogeneity awareness},
  author={Li, Dacheng and Wang, Hongyi and Xing, Eric and Zhang, Hao},
  journal={Advances in Neural Information Processing Systems},
  volume={35},
  pages={6630--6639},
  year={2022}
}

@inproceedings{um2024metis,
  title={Metis: Fast automatic distributed training on heterogeneous GPUs},
  author={Um, Taegeon and Oh, Byungsoo and Kang, Minyoung and Lee, Woo-Yeon and Kim, Goeun and Kim, Dongseob and Kim, Youngtaek and Muzzammil, Mohd and Jeon, Myeongjae},
  booktitle={2024 USENIX Annual Technical Conference (USENIX ATC 24)},
  pages={563--578},
  year={2024}
}

@inproceedings{yanhexiscale,
  title={HexiScale: Facilitating Large Language Model Training over Heterogeneous Hardware},
  author={Yan, Ran and Jiang, Youhe and Nie, Xiaonan and Fu, Fangcheng and Cui, Bin and Yuan, Binhang},
  booktitle={Ninth Conference on Machine Learning and Systems}
}

@inproceedings{hwang2024fasop,
  title={Fasop: Fast yet accurate automated search for optimal parallelization of transformers on heterogeneous gpu clusters},
  author={Hwang, Sunyeol and Lee, Eungyeong and Oh, Hongseok and Yi, Youngmin},
  booktitle={Proceedings of the 33rd International Symposium on High-Performance Parallel and Distributed Computing},
  pages={253--266},
  year={2024}
}

@article{chen2016training,
  title={Training deep nets with sublinear memory cost},
  author={Chen, Tianqi and Xu, Bing and Zhang, Chiyuan and Guestrin, Carlos},
  journal={arXiv preprint arXiv:1604.06174},
  year={2016}
}

@article{korthikanti2023reducing,
  title={Reducing activation recomputation in large transformer models},
  author={Korthikanti, Vijay Anand and Casper, Jared and Lym, Sangkug and McAfee, Lawrence and Andersch, Michael and Shoeybi, Mohammad and Catanzaro, Bryan},
  journal={Proceedings of Machine Learning and Systems},
  volume={5},
  pages={341--353},
  year={2023}
}

@article{huang2018gpipe,
  title={Gpipe: Efficient training of giant neural networks using pipeline parallelism},
  author={Huang, Yanping and Cheng, Youlong and Bapna, Ankur and Firat, Orhan and Chen, Mia Xu and Chen, Dehao and Lee, HyoukJoong and Ngiam, Jiquan and Le, Quoc V and Wu, Yonghui and others},
  journal={arXiv preprint arXiv:1811.06965},
  year={2018}
}

@inproceedings{jia2018exploring,
  title={Exploring hidden dimensions in parallelizing convolutional neural networks},
  author={Jia, Zhihao and Lin, Sina and Qi, Charles R and Aiken, Alex},
  booktitle={ICML},
  volume={2279},
  pages={2279--2288},
  year={2018}
}

@inproceedings{wang2019supporting,
  title={Supporting very large models using automatic dataflow graph partitioning},
  author={Wang, Minjie and Huang, Chien-chin and Li, Jinyang},
  booktitle={Proceedings of the Fourteenth EuroSys Conference 2019},
  pages={1--17},
  year={2019}
}

@inproceedings{song2019hypar,
  title={Hypar: Towards hybrid parallelism for deep learning accelerator array},
  author={Song, Linghao and Mao, Jiachen and Zhuo, Youwei and Qian, Xuehai and Li, Hai and Chen, Yiran},
  booktitle={2019 IEEE international symposium on high performance computer architecture (HPCA)},
  pages={56--68},
  year={2019},
  organization={IEEE}
}

@article{jia2019beyond,
  title={Beyond data and model parallelism for deep neural networks.},
  author={Jia, Zhihao and Zaharia, Matei and Aiken, Alex},
  journal={Proceedings of Machine Learning and Systems},
  volume={1},
  pages={1--13},
  year={2019}
}

@inproceedings{song2020accpar,
  title={Accpar: Tensor partitioning for heterogeneous deep learning accelerators},
  author={Song, Linghao and Chen, Fan and Zhuo, Youwei and Qian, Xuehai and Li, Hai and Chen, Yiran},
  booktitle={2020 IEEE International Symposium on High Performance Computer Architecture (HPCA)},
  pages={342--355},
  year={2020},
  organization={IEEE}
}

@article{cai2021tensoropt,
  title={Tensoropt: Exploring the tradeoffs in distributed dnn training with auto-parallelism},
  author={Cai, Zhenkun and Yan, Xiao and Ma, Kaihao and Wu, Yidi and Huang, Yuzhen and Cheng, James and Su, Teng and Yu, Fan},
  journal={IEEE Transactions on Parallel and Distributed Systems},
  volume={33},
  number={8},
  pages={1967--1981},
  year={2021},
  publisher={IEEE}
}

@inproceedings{zheng2022alpa,
  title={Alpa: Automating inter-and Intra-Operator parallelism for distributed deep learning},
  author={Zheng, Lianmin and Li, Zhuohan and Zhang, Hao and Zhuang, Yonghao and Chen, Zhifeng and Huang, Yanping and Wang, Yida and Xu, Yuanzhong and Zhuo, Danyang and Xing, Eric P and others},
  booktitle={16th USENIX Symposium on Operating Systems Design and Implementation (OSDI 22)},
  pages={559--578},
  year={2022}
}

@inproceedings{zhang2024hap,
  title={Hap: Spmd dnn training on heterogeneous gpu clusters with automated program synthesis},
  author={Zhang, Shiwei and Diao, Lansong and Wu, Chuan and Cao, Zongyan and Wang, Siyu and Lin, Wei},
  booktitle={Proceedings of the Nineteenth European Conference on Computer Systems},
  pages={524--541},
  year={2024}
}

@article{liu2023colossal,
  title={Colossal-auto: Unified automation of parallelization and activation checkpoint for large-scale models},
  author={Liu, Yuliang and Li, Shenggui and Fang, Jiarui and Shao, Yanjun and Yao, Boyuan and You, Yang},
  journal={arXiv preprint arXiv:2302.02599},
  year={2023}
}

@inproceedings{narayanan2021memory,
  title={Memory-efficient pipeline-parallel dnn training},
  author={Narayanan, Deepak and Phanishayee, Amar and Shi, Kaiyu and Chen, Xie and Zaharia, Matei},
  booktitle={International Conference on Machine Learning},
  pages={7937--7947},
  year={2021},
  organization={PMLR}
}

@inproceedings{fan2021dapple,
  title={DAPPLE: A pipelined data parallel approach for training large models},
  author={Fan, Shiqing and Rong, Yi and Meng, Chen and Cao, Zongyan and Wang, Siyu and Zheng, Zhen and Wu, Chuan and Long, Guoping and Yang, Jun and Xia, Lixue and others},
  booktitle={Proceedings of the 26th ACM SIGPLAN Symposium on Principles and Practice of Parallel Programming},
  pages={431--445},
  year={2021}
}

@article{miao2022galvatron,
  title={Galvatron: Efficient transformer training over multiple gpus using automatic parallelism},
  author={Miao, Xupeng and Wang, Yujie and Jiang, Youhe and Shi, Chunan and Nie, Xiaonan and Zhang, Hailin and Cui, Bin},
  journal={arXiv preprint arXiv:2211.13878},
  year={2022}
}

@inproceedings{liu2024aceso,
  title={Aceso: Efficient parallel DNN training through iterative bottleneck alleviation},
  author={Liu, Guodong and Miao, Youshan and Lin, Zhiqi and Shi, Xiaoxiang and Maleki, Saeed and Yang, Fan and Bao, Yungang and Wang, Sa},
  booktitle={Proceedings of the Nineteenth European Conference on Computer Systems},
  pages={163--181},
  year={2024}
}

@inproceedings{qi2026hetauto,
  title={HetAuto: Cross-Cluster Auto-Parallelism for Heterogeneous Distributed Training},
  author={Qi, Guicheng and Su, Junwei and Yang, Liqi and Li, Tao and Xie, Tingwen and Sun, Yerui and Xie, Yuchen and Wu, Chuan},
  booktitle={Proceedings of the 21st European Conference on Computer Systems},
  pages={759--779},
  year={2026}
}

@inproceedings{zhang2025poplar,
  title={Poplar: Efficient scaling of distributed DNN training on heterogeneous GPU clusters},
  author={Zhang, WenZheng and Hu, Yang and Shi, Jing and Bai, Xiaoying},
  booktitle={Proceedings of the AAAI Conference on Artificial Intelligence},
  volume={39},
  number={21},
  pages={22587--22595},
  year={2025}
}

@inproceedings{narayanan2019pipedream,
  title={PipeDream: Generalized pipeline parallelism for DNN training},
  author={Narayanan, Deepak and Harlap, Aaron and Phanishayee, Amar and Seshadri, Vivek and Devanur, Nikhil R and Ganger, Gregory R and Gibbons, Phillip B and Zaharia, Matei},
  booktitle={Proceedings of the 27th ACM symposium on operating systems principles},
  pages={1--15},
  year={2019}
}

@inproceedings{narayanan2021efficient,
  title={Efficient large-scale language model training on gpu clusters using megatron-lm},
  author={Narayanan, Deepak and Shoeybi, Mohammad and Casper, Jared and LeGresley, Patrick and Patwary, Mostofa and Korthikanti, Vijay and Vainbrand, Dmitri and Kashinkunti, Prethvi and Bernauer, Julie and Catanzaro, Bryan and others},
  booktitle={Proceedings of the international conference for high performance computing, networking, storage and analysis},
  pages={1--15},
  year={2021}
}

@inproceedings{chen2016xgboost,
  title={Xgboost: A scalable tree boosting system},
  author={Chen, Tianqi and Guestrin, Carlos},
  booktitle={Proceedings of the 22nd acm sigkdd international conference on knowledge discovery and data mining},
  pages={785--794},
  year={2016}
}

@article{snoek2012practical,
  title={Practical bayesian optimization of machine learning algorithms},
  author={Snoek, Jasper and Larochelle, Hugo and Adams, Ryan P},
  journal={Advances in neural information processing systems},
  volume={25},
  year={2012}
}

@article{touvron2023llama,
  title={Llama: Open and efficient foundation language models},
  author={Touvron, Hugo and Lavril, Thibaut and Izacard, Gautier and Martinet, Xavier and Lachaux, Marie-Anne and Lacroix, Timoth{\'e}e and Rozi{\`e}re, Baptiste and Goyal, Naman and Hambro, Eric and Azhar, Faisal and others},
  journal={arXiv preprint arXiv:2302.13971},
  year={2023}
}

@article{touvron2023llama2,
  title={Llama 2: Open foundation and fine-tuned chat models},
  author={Touvron, Hugo and Martin, Louis and Stone, Kevin and Albert, Peter and Almahairi, Amjad and Babaei, Yasmine and Bashlykov, Nikolay and Batra, Soumya and Bhargava, Prajjwal and Bhosale, Shruti and others},
  journal={arXiv preprint arXiv:2307.09288},
  year={2023}
}

@article{shoeybi2019megatron,
  title={Megatron-lm: Training multi-billion parameter language models using model parallelism},
  author={Shoeybi, Mohammad and Patwary, Mostofa and Puri, Raul and LeGresley, Patrick and Casper, Jared and Catanzaro, Bryan},
  journal={arXiv preprint arXiv:1909.08053},
  year={2019}
}

@inproceedings{li2021chimera,
  title={Chimera: efficiently training large-scale neural networks with bidirectional pipelines},
  author={Li, Shigang and Hoefler, Torsten},
  booktitle={Proceedings of the International Conference for High Performance Computing, Networking, Storage and Analysis},
  pages={1--14},
  year={2021}
}

@inproceedings{strati2025sailor,
  title={Sailor: Automating distributed training over dynamic, heterogeneous, and geo-distributed clusters},
  author={Strati, Foteini and Zhang, Zhendong and Manos, George and P{\'e}riz, Ixeia S{\'a}nchez and Hu, Qinghao and Chen, Tiancheng and Buzcu, Berk and Han, Song and Delgado, Pamela and Klimovic, Ana},
  booktitle={Proceedings of the ACM SIGOPS 31st Symposium on Operating Systems Principles},
  pages={204--220},
  year={2025}
}

@article{zong2025training,
  title={Training large models on heterogeneous and geo-distributed resource with constricted networks},
  author={Zong, Zan and Guo, Minkun and Zhai, Mingshu and Tang, Yinan and Li, Jianjiang and Zhai, Jidong},
  journal={Big Data Mining and Analytics},
  volume={8},
  number={4},
  pages={966--980},
  year={2025},
  publisher={TUP}
}

@article{li2025malleus,
  title={Malleus: Straggler-resilient hybrid parallel training of large-scale models via malleable data and model parallelization},
  author={Li, Haoyang and Fu, Fangcheng and Ge, Hao and Lin, Sheng and Wang, Xuanyu and Niu, Jiawen and Wang, Yujie and Zhang, Hailin and Nie, Xiaonan and Cui, Bin},
  journal={Proceedings of the ACM on Management of Data},
  volume={3},
  number={3},
  pages={1--28},
  year={2025},
  publisher={ACM New York, NY, USA}
}

@inproceedings{liang2026harp,
  title={HARP: Orchestrating Automated Parallel Training on Heterogeneous GPU Clusters},
  author={Liang, Antian and Zhao, Zhigang and Zhang, Kai and Shi, Xuri and Li, Chuantao and Wang, Chunxiao and He, Zhenying and Jing, Yinan and Wang, X Sean},
  booktitle={Proceedings of the 21st European Conference on Computer Systems},
  pages={1160--1180},
  year={2026}
}

@inproceedings{wang2026autohaap,
  title={AutoHAAP: Automated Heterogeneity-Aware Asymmetric Partitioning for LLM Training},
  author={Wang, Yuanyuan and Tang, Nana and Wang, Yuyang and Pan, Shu and Yu, Dingding and Wang, Zeyue and Sun, Mou and Fu, Kejie and Wang, Fangyu and Chen, Yunchuan and others},
  booktitle={2026 IEEE International Symposium on High Performance Computer Architecture (HPCA)},
  pages={1--17},
  year={2026},
  organization={IEEE}
}

@inproceedings{zhu2025mist,
  title={Mist: Efficient Distributed Training of Large Language Models via Memory-Parallelism Co-Optimization},
  author={Zhu, Zhanda and Giannoula, Christina and Andoorveedu, Muralidhar and Su, Qidong and Mangalam, Karttikeya and Zheng, Bojian and Pekhimenko, Gennady},
  booktitle={Proceedings of the Twentieth European Conference on Computer Systems},
  pages={1298--1316},
  year={2025}
}

@article{li2025oases,
  title={Oases: Efficient Large-Scale Model Training on Commodity Servers via Overlapped and Automated Tensor Model Parallelism},
  author={Li, Shengwei and Lai, Zhiquan and Li, Dongsheng and Hao, Yanqi and Liu, Weijie and Ge, Keshi and Deng, Xiaoge and Lu, Kai},
  journal={IEEE Transactions on Parallel and Distributed Systems},
  year={2025},
  publisher={IEEE}
}

@article{guo2025adaptis,
  title={OctoPipe: Reducing Pipeline Bubbles for Heterogeneous Models via Co-Optimizing Partitioning, Placement, and Scheduling},
  author={Guo, Jihu and Ma, Tenghui and Gao, Wei and Sun, Peng and Li, Jiaxing and Chen, Xun and Jin, Yuyang and Lin, Dahua},
  journal={arXiv preprint arXiv:2509.23722},
  year={2025}
}

@inproceedings{ou2025neural,
  title={A Neural Network-Based Pipeline Parallel Strategy Solver for Heterogeneous Environments},
  author={Ou, Jie and Guo, Jinyu and Chen, Yueming and Jiang, Shuaihong and Xue, Ruini and Tian, Wenhong},
  booktitle={2025 International Joint Conference on Neural Networks (IJCNN)},
  pages={1--9},
  year={2025},
  organization={IEEE}
}

@inproceedings{ou2025efficient,
  title={Efficient and Automatic 3D Parallelism Strategies Search via Contrastive Reinforcement Learning Pretrained Neural Networks},
  author={Ou, Jie and Li, Xiaowang and Chen, Yueming and Qian, Jiahong and Su, Teng and Tian, Wenhong},
  booktitle={2025 IEEE 12th International Conference on Data Science and Advanced Analytics (DSAA)},
  pages={1--10},
  year={2025},
  organization={IEEE}
}

@article{shong2026autoscout,
  title={AutoScout: Structured Optimization for Automating ML System Configuration},
  author={Shong, Jimmy and Ding, Yuhan and Jiang, Yihan and Jing, Liheng and Chen, Haonan and Zhang, Gaokai and Akella, Aditya and Lai, Fan},
  journal={arXiv preprint arXiv:2603.11603},
  year={2026}
}

@article{wang2024pro,
  title={Pro-Prophet: A Systematic Load Balancing Method for Efficient Parallel Training of Large-scale MoE Models},
  author={Wang, Wei and Lai, Zhiquan and Li, Shengwei and Liu, Weijie and Ge, Keshi and Shen, Ao and Su, Huayou and Li, Dongsheng},
  journal={arXiv preprint arXiv:2411.10003},
  year={2024}
}

@inproceedings{yun2026hpmd,
  title={HPMD: Enabling Hybrid Parallelism with Multi-Dimensional Adaptive DNN Training},
  author={Yun, Gyeongchan and Choi, Young-ri},
  booktitle={Proceedings of the 40th ACM International Conference on Supercomputing},
  pages={1027--1039},
  year={2026}
}

@inproceedings{wu2025rethinking,
  title={Rethinking Dynamic Networks and Heterogeneous Computing with Automatic Parallelization},
  author={Wu, Ruilong and Li, Xinjiao and Wang, Yisu and Chen, Xinyu and Kutscher, Dirk},
  booktitle={Proceedings of the 9th Asia-Pacific Workshop on Networking},
  pages={164--171},
  year={2025}
}

\vskip1pc

\bio{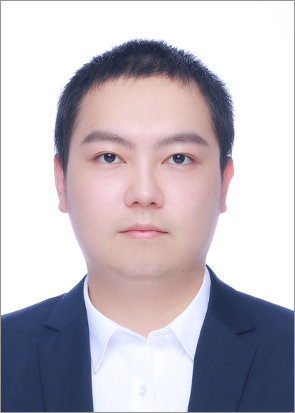}
Lin Meng received the B.S. degree from Southeast University, Jiangsu, China in 2020. He is pursuing his Ph.D. degree at the Institute of Computing Technology (ICT), Chinese Academy of Sciences. His research interests include distributed machine learning systems and communication optimizations.
\endbio

\vskip3pc

\bio{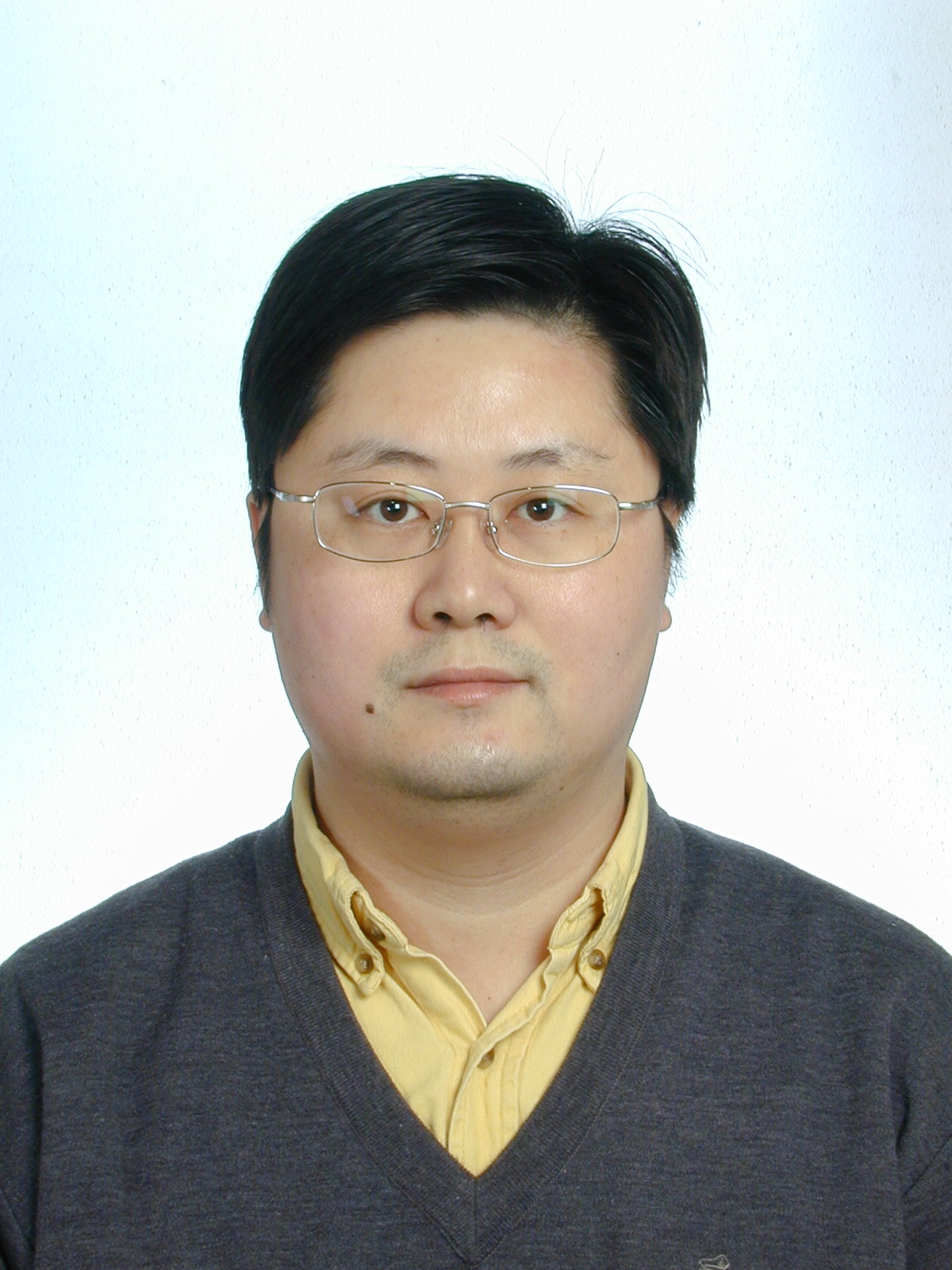}
Yuzhong Sun received the PhD degree in computer engineering from the Institute of Computing Technology (ICT), Chinese Academy of Sciences. He is a professor at ICT, China. His research interests focus on distributed system software and computing/programming for model training and inference.
\endbio

\vskip3pc

\bio{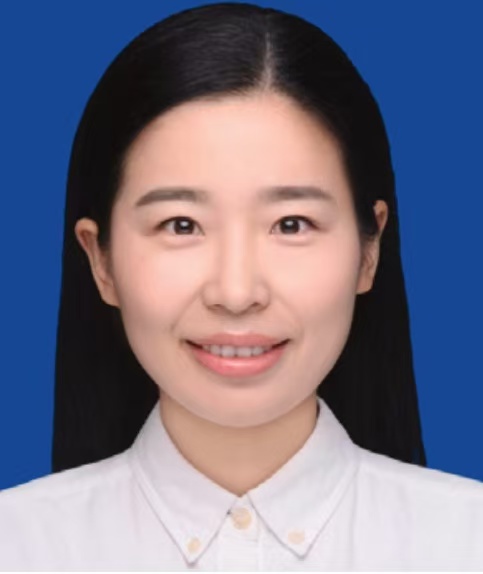}
Jie Zhu received the Ph.D degree in applied computer science from the School of Computer Science and Engineering, Southeast University, Nanjing, in 2011. She joined Nanjing University of Post \& Telecommunication in 2014, where she is currently an Associate Professor with the School of Computer Science. Her research interests include Machine Scheduling, Project Scheduling, Workflow Optimization and Cloud Computing.
\endbio

\end{document}